\newif\ifdraft\drafttrue

\documentclass[acmsmall,screen]{acmart}

\setcopyright{none}
\usepackage{xcolor}
\usepackage{subcaption}
\usepackage{wrapfig}
\usepackage{graphicx}
\usepackage{array}
\usepackage{longtable}
\usepackage{tikz}
\usetikzlibrary{patterns}

\definecolor{chart_quickcheck}{HTML}{1B7837}
\definecolor{chart_hedgehog}{HTML}{2166AC}
\definecolor{chart_falsify}{HTML}{B35806}
\definecolor{bucketbg}{RGB}{240,240,240}
\newcommand{\explaincolor}[2]{{\raisebox{-.3ex}{\color{#1}\rule{1.2em}{.8em}}}~= #2}
 \newcommand{\explaincolorhatched}[2]{{\raisebox{-.3ex}{\begin{tikzpicture}\fill[#1] (0,0) rectangle (1.2em,.8em); \fill[pattern=north east lines, pattern color=white] (0,0) rectangle (1.2em,.8em);\end{tikzpicture}}}~= #2}
\definecolor{dkgreen}{rgb}{0,0.6,0}
\definecolor{ltblue}{rgb}{0,0.4,0.4}
\definecolor{dkviolet}{rgb}{0.3,0,0.5}

\usepackage{listings}

\lstdefinestyle{haskellstyle}{
	language=Haskell,
	escapechar=@,
	basicstyle=\ttfamily\small,
	showstringspaces=false,
	columns=[l]flexible,
	identifierstyle={\ttfamily\color{black}},
	keywordstyle={\ttfamily\color{dkviolet}},
	keywordstyle=[2]{\ttfamily\color{dkblue}},
	keywordstyle=[3]{\ttfamily\color{ltblue}},
	stringstyle={\ttfamily\color{dkred}},
	commentstyle={\ttfamily\upshape\color{dkgreen}},
	morekeywords=[2]{case,of,let,in,where,if,then,else,do},
	morekeywords=[3]{IO,Int,Integer,Bool,Char,Maybe,Either},
    deletekeywords={String, StdGen, newStdGen},
}
\lstnewenvironment{hask}{\lstset{style=haskellstyle}}{}
\newcommand{\HC}{\lstinline[style=haskellstyle, basicstyle=\ttfamily\small]}

\begin{document}

\title{Evaluating Shrinking (Experience Report)}

\author{Alperen Keles}
\email{akeles@umd.edu}
\affiliation{%
	\institution{University of Maryland}
	\city{College Park}
	\country{USA}
}

\author{George Miao}
\email{gmiao@umd.edu}
\affiliation{%
	\institution{University of Maryland}
	\city{College Park}
	\country{USA}
}

\author{Leonidas Lampropoulos}
\email{leonidas@umd.edu}
\affiliation{%
	\institution{University of Maryland}
	\city{College Park}
	\country{USA}
}

\begin{abstract}
  Property-based testing frameworks rely on \emph{shrinking} to turn noisy random
  failures into counterexamples that developers can debug. Although bug-finding
  performance is routinely measured, shrinking itself is rarely evaluated
  quantitatively. We present an experience report on evaluating shrinking across
  three Haskell frameworks: QuickCheck, Hedgehog, and Falsify. The comparison spans
  four ETNA workloads and several generator families, including type-based,
  API-based, and correct-by-construction generators. We measure both
  effectiveness, using tree edit distance to a ground-truth minimum found by
  exhaustive search, and cost, using shrink time and time per unit of shrinking
  progress. Across these workloads, QuickCheck's structural shrinking is usually
  faster and remains competitive on final counterexample quality; integrated
  shrinking does not by itself guarantee a performance or effectiveness advantage.
  We discuss what these results imply for future evaluations and designs of
  shrinking algorithms.
\end{abstract}


\maketitle

\section{Introduction}

Property-based testing is a well-established paradigm for gaining
confidence in the correctness of programs. Popularized by Haskell's
QuickCheck~\cite{ClaessenH00}, there has been an explosion of
property-based testing frameworks in the last 25 years across
languages~\cite{Hypothesis,ScalaCheck,QCheckOCaml,quickcheck-rs}.

Across all these frameworks, some key characteristics stay the same.
First, users must provide {\em properties} of the system under
test in the form of executable predicates over universally quantified
inputs. For example, to test the correctness of a binary search tree
implementation using Haskell's QuickCheck, one could write the following
property that dictates that if a given binary search tree is valid, i.e. that
it satisfies the standard search tree invariant, then a correct implementation
of \HC{insert} will preserve this invariant:

\begin{hask}
		prop_insert :: Tree -> Key -> Value -> Maybe Bool
		prop_insert t k v = isBST t ==> isBST (insert t k v)
\end{hask}

To test this property, users must also specify how to generate test
inputs.  There are plenty of ways to achieve that, and even more
literature on that matter: one can generate trees based on the type of
the predicate and filter out those that are invalid~\cite{ClaessenH00};
one can write a specialized generator that only produces valid
trees~\cite{Hughes2019HowTS}; one could devise automatic procedures that
derive such generators automatically~\cite{GeneratingGoodGenerators,Luck}. In all those
cases, the framework takes such a generator plus the property above as
an input and embarks on a straightforward generate-and-test loop that
repeatedly generates inputs, randomly exploring the large space of
possible inputs until a counterexample is found.

However, the very same randomness that tames the exponentially large
search space also means that most of the time the first counterexample
found is unusable for debugging: it simply contains too much noise.
For example, leveraging a random tree generator to test the
\HC{prop_insert} property above, often yields counterexamples like the
following:

\begin{lstlisting}[basicstyle=\ttfamily\scriptsize, breaklines=true,
	breakatwhitespace=true, breakautoindent=false, breakindent=0pt,
	columns=fullflexible, xleftmargin=1.2em,
	aboveskip=0.8em, belowskip=0.8em]
((T (T (T (E) -684 128 (E)) -563 533 (T (E) -552 252 (T (T (E) -479 27 (E)) -412 -910 (T (T (E) -395 79 (T (E) -349 -332 (E))) -330 779 (E))))) -253 -554 (T (T (T (T (T (T (E) -252 -550 (T (T (T (E) -231 -330 (E)) -222 -19 (E)) -175 279 (T (E) -113 136 (T (E) -111 170 (E))))) -48 30 (E)) -25 238 (E)) 76 632 (T (T (T (E) 85 -717 (T (E) 90 -781 (E))) 161 167 (E)) 192 -295 (T (T (T (E) 200 -19 (E)) 256 -286 (E)) 285 -808 (T (E) 306 156 (E))))) 401 260 (T (E) 402 95 (E))) 460 661 (T (T (E) 545 -389 (E)) 743 60 (E)))), -25, 5)
\end{lstlisting}

\noindent Such an input is difficult to use for debugging: of the 31
nodes in the tree, only one is needed to trigger the fault. If only we
could devise a method to simplify this input to the core of the bug, we
could present the user with a simpler counterexample such as
\HC{(T (E) -25 0 (E), -25, 1)}, from which the bug is immediately
apparent.

The solution to this problem is counterexample minimization, also
known as {\em shrinking}. QuickCheck's approach to shrinking is a
second straightforward shrink-and-test loop: users provide a shrinking
function---i.e., a function that given a counterexample produces a
list of smaller variations of it---and the framework repeatedly tries
all such variations until one of them is found to also be a
counterexample, repeating this process until a (local) minimum is
found. This particular approach is called {\em external} shrinking;
the shrinker is written independently from the generator and focuses
on minimizing the generated structure. This independence poses a
problem that valid counterexamples (produced by hand-crafted generators
that produce inputs valid-by-construction) are minimized into invalid
candidates, spending the testing budget in sifting through the invalid
cases.

It would be nice if we could have correct-by-construction shrinkers too,
and that is precisely the problem solved by \emph{integrated}, or
\emph{internal}, shrinking. Python's Hypothesis~\cite{Hypothesis} exemplifies integrated
shrinking as an alternative path, where instead of independent
shrinkers, the generators themselves are used in shrinking. That is
achieved by shrinking the randomness coming into the generators and
then replaying them, rather than shrinking the structures produced,
usually by representing this randomness as a \emph{randomness buffer}
or a \emph{choice sequence} to maintain some structural correspondence
with those outputs.  Variations of this approach are found in other
Haskell property-based testing frameworks, most prominently
Hedgehog~\cite{Hedgehog} and Falsify~\cite{falsify}.

How do these approaches compare to each other? Integrated shrinking is
quite appealing in theory: it relieves the users of the responsibility
to implement shrinkers, automatically giving rise to
correct-by-construction shrinkers instead of requiring users to
program them themselves. However, as noted in the study of the
Hypothesis reducer~\cite{HypothesisShrinking}, external shrinkers can be
more \emph{effective} in finding smaller counterexamples than internal
shrinkers. For example, code generators for testing compilers might
produce test inputs with a particular boilerplate structure which will
always be present with internal shrinking, regardless of whether it's
necessary. The performance implications of the shrinking choice is
also a key concern: given a fixed testing budget, overhead from using
an internal-shrinking-based approach can mean that the testing campaign will
be able to cover a smaller part of the search space. Ideally, the
users should have an informed view of the trade-offs present in
selecting between different approaches.

The PBT literature, however, overwhelmingly focuses on evaluating bug-finding performance
with little quantitative evaluation of the shrinking process. There are a number
of metrics in comparing bug-finding performances for different generation strategies
such as measuring code coverage~\cite{FuzzChick} or running mutations tests~\cite{TestingNIjfp};
ETNA~\cite{ETNA} evaluation platform provides the users with a diverse set of workloads
as well as measurement tools for bug-finding performance, but it provides no support
in evaluating shrinking.
This surprising lack of evaluation of shrinking approaches is exactly the motivation
for this paper: we set out to develop a comparative understanding of existing
minimization approaches. In order to keep the comparison fair in terms of usability, we have
opted for minimum effort generic shrinkers in all our evaluations.

Concretely, we offer the following contributions:

\vspace{-2mm}

\begin{itemize}
	\item We define quantitative metrics for shrinking effectiveness and cost,
			including tree edit distance to a ground-truth minimum and time per unit
			of shrinking progress.
	\item We extend the ETNA PBT evaluation platform with shrinking measurements
			and apply it to QuickCheck, Hedgehog, and Falsify across four Haskell
			workloads.
	\item We report how shrinking behavior changes across type-based,
			API-based, and correct-by-construction generators, and
			discuss the implications for PBT library design.
\end{itemize}

\section{Background: Understanding Shrinking}
\label{sec:technical}

Shrinking is a conceptually simple problem. The shrinking
algorithm generates smaller candidates, checks whether they still reproduce the failure,
and repeats until no smaller failing input is found. The search typically reaches
a local minimum rather than a guaranteed global minimum, so the structure of the
candidate space matters.

Researchers have proposed various implementations of this general
pattern, which can be grouped under two primary umbrellas: the first
is what is called ``external'', ``type-based'', or ``structural''
shrinking, which involves explicit functions operating directly on the
structure of the counterexamples. The second is what has been
historically called ``internal'' or ``integrated'' shrinking, as
shrinking behavior is integrated into generators.
This section briefly presents the respective techniques used in the
Haskell PBT frameworks we evaluate.

\subsection{Structural Shrinking---QuickCheck}
\label{subsec:structural}

In structural shrinking, the user provides a \emph{shrinker}, a pure function
\HC{a -> [a]} that maps a value to a list of \emph{candidate
	smaller values}. When a test fails, the framework searches those candidates for
a new failing input, repeating until no candidate reduces further. The shrinker
operates directly on the structure of the value and is entirely independent of
the generator. Take QuickCheck's implementation as an example, the user-facing API
is the \texttt{Arbitrary} typeclass:

\begin{hask}
	class Arbitrary a where
	    arbitrary :: Gen a
	    shrink :: a -> [a]
\end{hask}

\noindent The \HC{arbitrary} and \HC{shrink} methods are decoupled. The shrinker has no
access to the \HC{Gen} monad and no knowledge of how the value was produced.
This means any invariants encoded in the generator must be re-enforced
independently in \HC{shrink}---if the shrinker produces a candidate that
violates a precondition, the property may fail for the wrong reason.

QuickCheck implements this typeclass for basic Haskell types, such as integers (which
shrink to some smaller number) or lists (which can shrink to a sublist or to a list
where one of its elements has recursively been shrunk). Still, users must provide
this implementation for any user-defined datatypes, or any types whose shrinking
behavior they want to override.

\paragraph{The Search Loop}

Internally, QuickCheck represents the shrink space as a lazy rose tree~\cite{ClaessenH00}:

\begin{hask}
	data Rose a = MkRose a [Rose a]
\end{hask}

\noindent
Each \HC{MkRose v cs} node holds a test result \HC{v} paired with the shrink
candidates of \HC{v}, each rooted in their own subtree. The tree is built
lazily, with only the subtree actually visited ever being forced.

Once a failure is found, QuickCheck searches the tree with a greedy left-to-right
depth-first traversal. At each step, if the first candidate \HC{t} still fails the
property,
descend into its children \HC{ts'}; otherwise advance to the
next sibling \HC{ts}. The algorithm never backtracks, so the result is a
\emph{local} minimum. This makes the ordering of the shrink list significant, since the
greedy search locks in the first improvement it finds, front-loading the list with
globally small values gives the algorithm its best chance of reaching the global
minimum.

\subsection{SampleTrees---Falsify}
\label{subsec:sampletrees}

In integrated shrinking, generators carry their shrinking behavior implicitly.
Instead of a separate shrinker function, the generator runs on an
internal randomness buffer, and shrinking works by shrinking the buffer
with smaller values and re-running the same generator. Falsify implements this
approach with \emph{SampleTrees}, a lazy tree structure representing both the
original random samples and all possible shrunk variants.

The user-facing API
is simply a generator monad:

\begin{hask}
    newtype Gen a = Gen { runGen :: SampleTree -> (a, [SampleTree]) }
\end{hask}

\noindent The \HC{Gen} monad consumes a \HC{SampleTree}, a lazy binary tree
of \HC{Word64} samples, and returns both the generated value and a list of
\emph{shrunk sample trees}. Each shrunk tree represents a candidate shrink of
the input: it replaces one or more samples with smaller values. When a
property fails, Falsify re-runs the generator on each shrunk tree until no
candidate produces a smaller failing input.










\subsection{Shrink Tree---Hedgehog}

Hedgehog also uses integrated shrinking, but exposes it through a
generator that directly produces a shrink tree. Internally, a generator
is a function of the current size and random seed:

\begin{hask}
    type Gen = GenT Identity

    newtype GenT m a = GenT {
        unGenT :: Size -> Seed -> TreeT (MaybeT m) a
    }
\end{hask}

\noindent Running a \HC{Gen} produces a lazy rose tree, where the root
is the generated test case and children are the shrink candidates.
Discarded inputs are represented by the \HC{MaybeT} layer. As
with Falsify, the user writes only a generator; the shrink behavior is
attached to the choices made by that generator instead of a separate function.







\section{Evaluation}
\label{sec:eval}

In our evaluation, we ask two questions, (1) how \emph{effective} is the shrinker
via measuring the distance to the truly minimal input, and (2) how \emph{performant} is the
shrinker via measuring the time it takes to shrink. Concretely, we measure the following metrics
for answering our questions:

\begin{itemize}
	\item \textbf{Minimal counterexample discovery:} Properties can tested with
	      many different techniques for supplying inputs, from manual user-written inputs to our usual
	      random input generation to symbolic execution to exhaustive enumeration~\cite{RuncimanNL08,Matela2017ToolsFD,CombinatorialEnum}.
		  We leverage an enumerative property-based testing strategy, LeanCheck~\cite{Matela2017ToolsFD},
		  to search for the smallest discovered input that constitutes a bug. We use these
		  LeanCheck results as ground-truth minima for the tasks where exhaustive search finds the bug.
	\item \textbf{Effectiveness of shrinking strategy:} Regardless of the original input, an effective
	      shrinker should move the reported counterexample closer to a minimal failing input. We use the tree edit distance~\cite{zhang1989simple}
	      computed via zss package in Python~\cite{zssPythonGithub} between the ground truth minimal counterexample and
		  the shrinker output to measure effectiveness.
	\item \textbf{Performance of shrinking strategy:} We measure shrinking time and time per unit of
	      tree-edit-distance reduction to understand how much time the shrinker spends per unit of progress
	      instead of just looking at the absolute numbers. We measure different shrinker parameter choices in
	      the libraries to understand their effects.
	\item \textbf{Cost of shrinking strategy:} There is an inherent cost to the shrinking strategies themselves
		  that affects the performance of the bug-finding stage of the testing process. We measure the bug-finding
		  capabilities of different frameworks as well as the overhead of enabling the shrinking.
	\item \textbf{Stability of shrinking strategies:} It is possible to write different generators for the same
		  input space. We have briefly mentioned how it is possible to write correct-by-construction generators that
		  target the valid subset of the input space instead of generators that samples the entire domain, but we can
		  also have different styles of correct-by-construction generators too, such as API-based generation,
		  which instead of generating the structure itself, generates operations that mutate it. We measure
		  the difference in shrinking performance and effectiveness in the context of different generation strategies
		  in our evaluation.
\end{itemize}

\subsection{Background: ETNA}
\label{sec:background}

ETNA~\cite{etna-icfp23} is an evaluation and analysis platform for Property-Based Testing frameworks.
It hosts \emph{workloads}, programs with injected mutations constituting bugs, as well
as properties violated by these mutations, with each property-mutation pair that can
be violated serving as a {\em task}.
These mutations are hand-written providing a ground truth evaluation, following a 
Magma~\cite{Hazimeh:2020:Magma} style historical bug injection methodology across its workloads.
Workloads are available in multiple languages, including Haskell and
its diverse PBT landscape. The Haskell-specific part of ETNA includes 4 workloads,
Binary-Search Tree with 53 tasks, Red-Black Tree with 58 tasks,
Simply-Typed Lambda Calculus with 20 tasks, and System $F_{<:}$ with
36 tasks.
The workloads are also accessible over a simple command-line interface that allows for
creating and running experiments; adding, inspecting and manipulating workloads
and tests; and generating reports on the collected results.

Testing strategies from different libraries are embedded within each workload.
The Haskell workloads originally had QuickCheck~\cite{ClaessenH00}, SmallCheck~\cite{RuncimanNL08}, LeanCheck~\cite{leancheck}; we extended them with Hedgehog~\cite{Hedgehog} and Falsify~\cite{falsify} for our evaluation.

\begin{figure}[h!]
	\centering
	\resizebox{\linewidth}{!}{%
	\begin{tikzpicture}[x=2.2cm, font=\small]
		\foreach \i/\lab in {%
			0/{<\,0.1\,s}, 1/{0.1--1\,s}, 2/{1--10\,s},
			3/{10--60\,s}, 4/{60--360\,s}, 5/{unsolved}}
			\node[anchor=south, font=\bfseries\footnotesize]
				at (\i+0.5, 0.1) {\lab};
		\foreach \r/\C/\name in {%
			0/chart_quickcheck/QuickCheck,
			1/chart_hedgehog/Hedgehog,
			2/chart_falsify/Falsify} {
			\pgfmathsetmacro{\ytop}{-\r*0.74}
			\pgfmathsetmacro{\ybot}{\ytop-0.62}
			\node[anchor=east, font=\bfseries\footnotesize]
				at (-0.1, {(\ytop+\ybot)/2}) {\name};
			\foreach \i in {0,...,5} {
				\pgfmathtruncatemacro{\p}{100-20*\i}
				\filldraw[fill=\C!\p!bucketbg, draw=white, line width=0.6pt]
					(\i, \ybot) rectangle (\i+1, \ytop);
			}
		}
	\end{tikzpicture}%
	}
	\caption{ETNA-style bucket-chart key}
	\label{fig:etna-legend}
\end{figure}
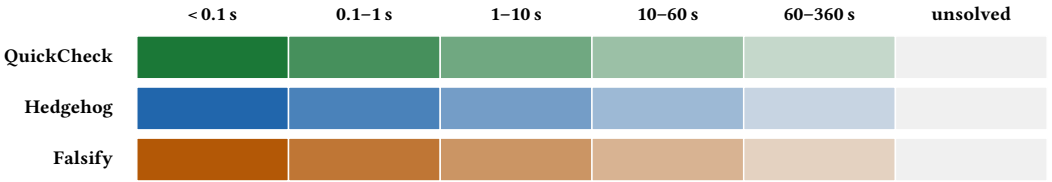

ETNA reports bug-finding performance using \emph{bucket charts}. Each
strategy is drawn as a single horizontal bar, partitioned into segments
whose widths are proportional to the number of \emph{tasks} that fall
into each time bucket. A task is placed in a bucket by the median
time-to-failure across its trials; tasks for which any trial failed to
find the bug within the timeout are counted as \emph{unsolved}. Each
framework keeps a fixed hue---green for QuickCheck, blue for Hedgehog,
orange for Falsify---and its segments are shaded from that full hue
(solved fastest) to near-white (unsolved), so a bar dominated by
saturated color denotes a strategy that finds bugs quickly.
Figure~\ref{fig:etna-legend} shows the bucket key used throughout this
paper; unlike the original ETNA experiments, we use a 360-second timeout and
therefore an additional \texttt{60--360\,s} bucket.

\subsection{Experiments}
\label{sec:experiments}

In our evaluation, we measure bug-finding time (time spent before a failing execution) and shrinking time (time spent after a failing execution) as performance metrics.
The libraries provide shrinking budget controllers, albeit not very well documented, that we use to turn off shrinking as well as
tune it with different parameters to measure its effect.

For BST and RBT, we used three different generators that represent the
most common generators in the literature: a naive, type-based
generator that uses rejection sampling for discarding invalid trees; a
correct-by-construction generator that directly generates valid trees;
and a second correct-by-construction generator that generates
key-value pairs that are inserted to the trees; we will refer to the
latter as an API-based generator, as it leverages the API of the
structure under test to generate instances that satisfy its
invariants. In all three frameworks--QuickCheck, Hedgehog and
Falsify--we tried to create the same generator design as closely as
each API allowed. Specifically for the correct-by-construction BST
generators (as can be seen in Appendix~\ref{app:bst-generators}), we
created additional idiomatic variants for Hedgehog (that
uses \texttt{HH.recursive}) and Falsify (that
mimics \texttt{Falsify.Generator.bst}) and evaluated the effects of
the changes. For STLC and $F_{<:}$, we ported the existing naive and
correct-by-construction QuickCheck generators. A notable usability
advantage of integrated shrinking is that it relieves the user of the
requirement to write shrinkers~\cite{falsify,HypothesisShrinking}. In order
to keep the comparison fair, we used the \texttt{genericShrink} from
\texttt{generic-random}~\cite{genericRandom} library that automatically
derives shrinkers for Haskell types.

\subsubsection{Comparison of Bug-Finding Performances}

\begin{figure}[h!]
	\centering
	\begin{subfigure}[t]{.33\linewidth}
		\centering
		\includegraphics[width=\linewidth]{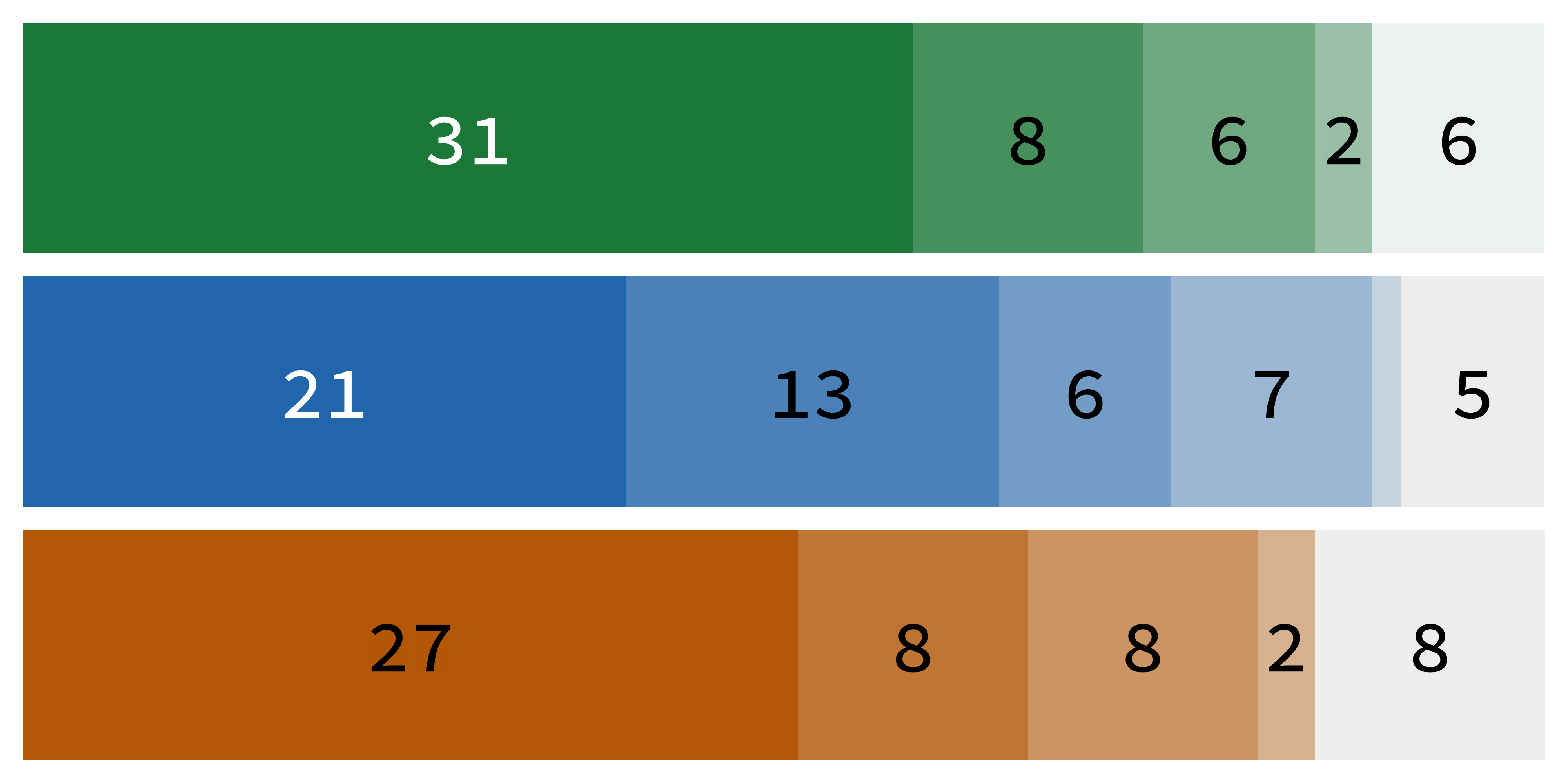}
		\caption{Type-based generation}
		\label{fig:bucket_bst_vanilla}
	\end{subfigure}\hfill
	\begin{subfigure}[t]{.33\linewidth}
		\centering
		\includegraphics[width=\linewidth]{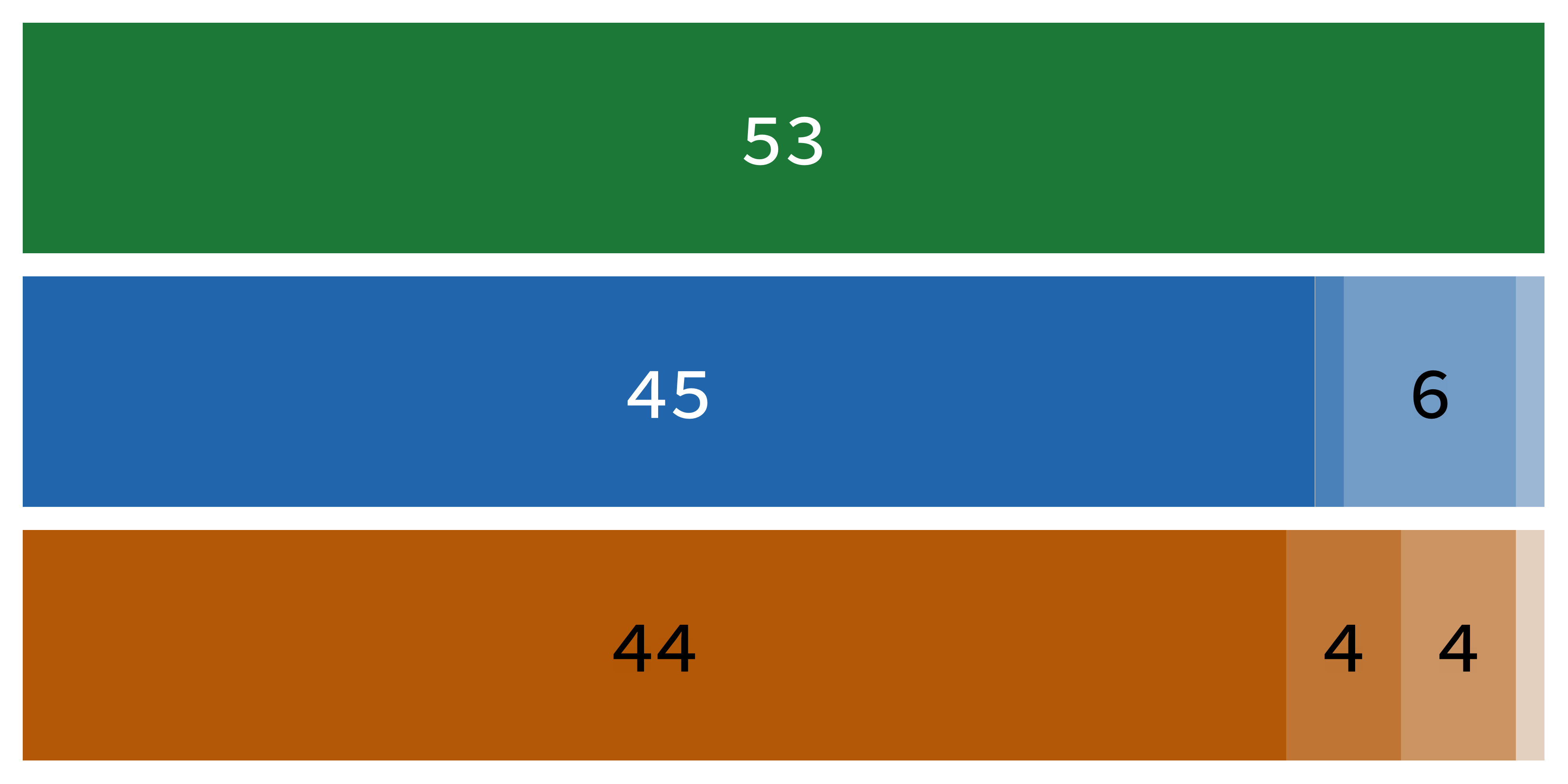}
		\caption{API}
		\label{fig:bucket_bst_qbe}
	\end{subfigure}
	\begin{subfigure}[t]{.33\linewidth}
		\centering
		\includegraphics[width=\linewidth]{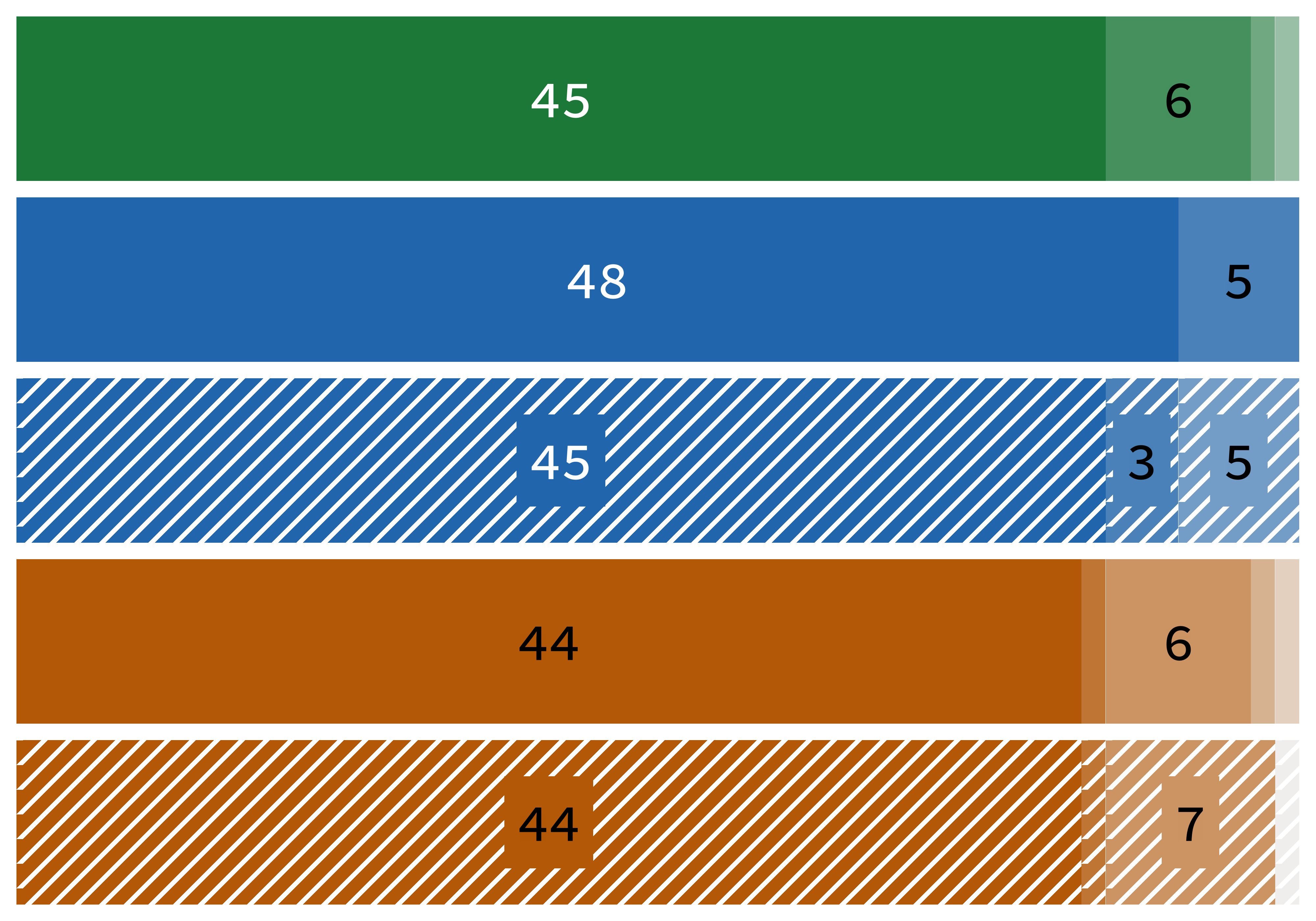}
		\caption{Correct-by-construction}
		\label{fig:bucket_bst_cbc}
	\end{subfigure}
	\caption{Bug-finding bucket charts on BST \\
		\explaincolor{chart_quickcheck}{QuickCheck},
		\explaincolor{chart_hedgehog}{Hedgehog},
		\explaincolorhatched{chart_hedgehog}{Idiomatic Hedgehog},\\
		\explaincolor{chart_falsify}{Falsify},
		\explaincolorhatched{chart_falsify}{Idiomatic Falsify}}
	\label{fig:bucket_bst}
\end{figure}

The BST results in Figure~\ref{fig:bucket_bst} match the intuitive expectation that 
correct-by-construction generators largely eliminate the wasted effort of naive type-based
generation, which spends substantial effort producing invalid trees. In the bucket charts,
almost all API-based and correct-by-construction generators solve every BST task, while the type-based
generators fail several tasks (\texttt{not found} bucket); API-based QuickCheck generator has the strongest
bug-finding profile, placing all tasks in the fastest bucket.

Across libraries, QuickCheck is significantly faster for bug-finding in both the type-based
and API-based settings. We model the experiments as a paired repeated-measures
design following Dem\v{s}ar~\cite{demsar2006statistical}: each task is a data set, trial-level
measurements are collapsed to per-task medians, and we run a Friedman test followed by
Holm-corrected Wilcoxon signed-rank post-hoc tests. For BST bug-finding time, the type-based
and API-based comparisons are statistically significant ($p < 0.001$):
QuickCheck has lower bug-finding times than both Hedgehog and Falsify, and Falsify also has lower bug-finding times than Hedgehog in
the type-based setting. In contrast, the correct-by-construction generators are statistically
indistinguishable for bug-finding time among successfully solved tasks.\footnote{To avoid assigning arbitrary penalties, the
paired statistical comparisons exclude \texttt{not found} cases and are therefore computed only
on tasks with successful failed trials for all compared strategies. This can differ from the
visual impression of the bucket charts, where \texttt{not found} is shown explicitly.}

\begin{figure}[h!]
	\centering
	\begin{subfigure}[t]{.30\linewidth}
		\centering
		\includegraphics[width=\linewidth]{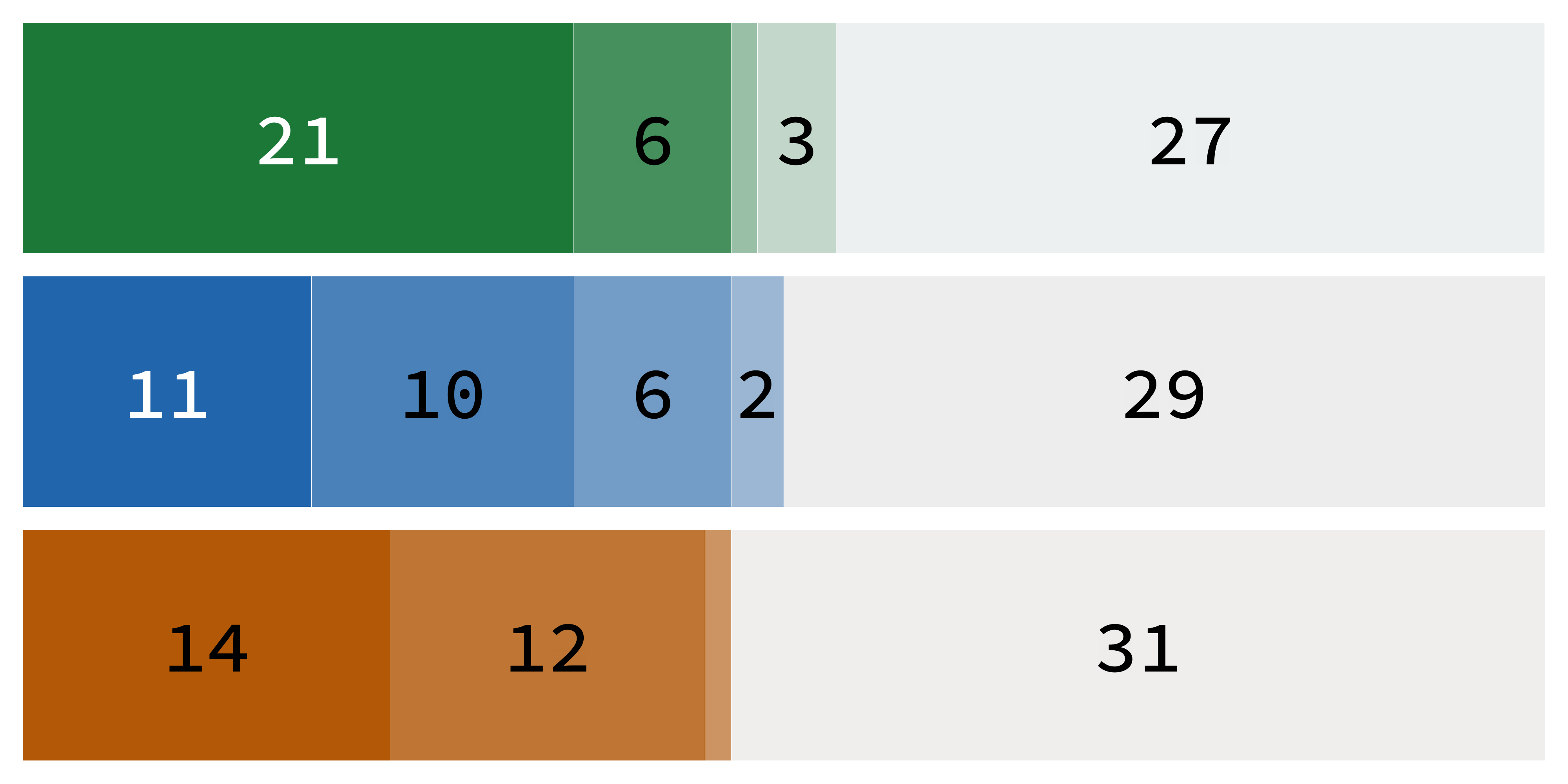}
		\caption{Type-based generation}
		\label{fig:bucket_rbt_vanilla}
	\end{subfigure}
	\begin{subfigure}[t]{.30\linewidth}
		\centering
		\includegraphics[width=\linewidth]{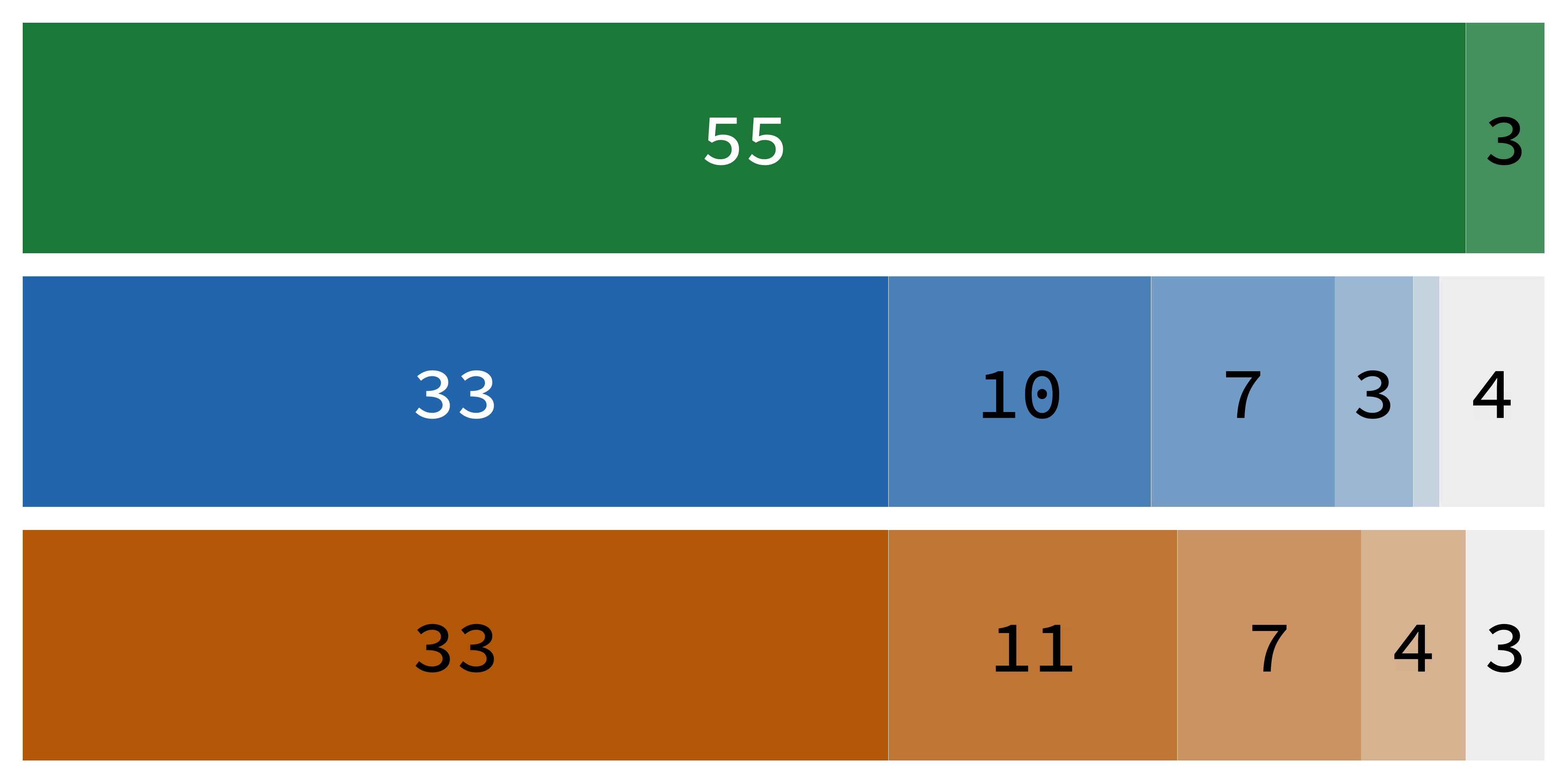}
		\caption{API}
		\label{fig:bucket_rbt_qbe}
	\end{subfigure}
	\begin{subfigure}[t]{.30\linewidth}
		\centering
		\includegraphics[width=\linewidth]{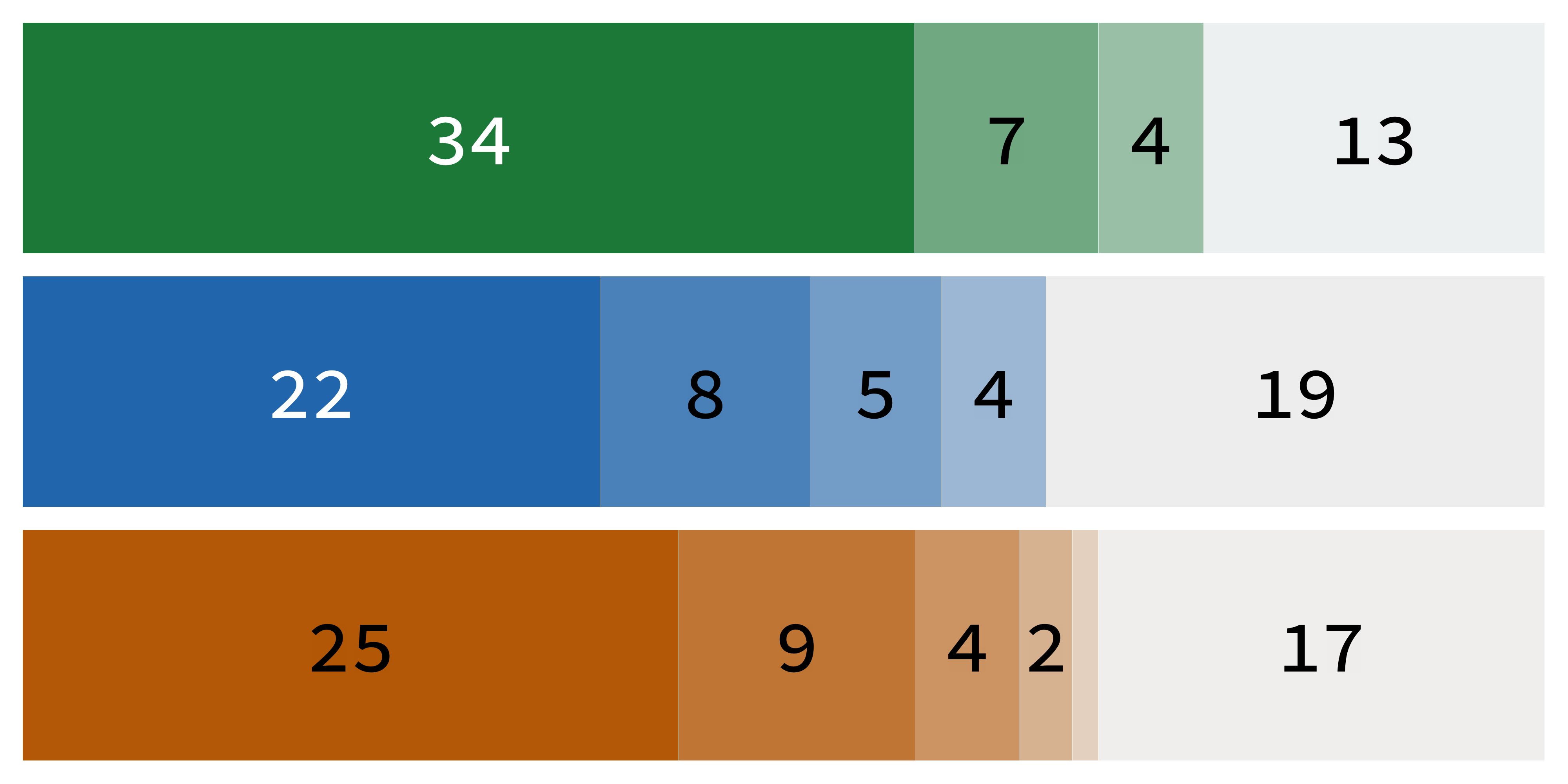}
		\caption{Correct-by-construction}
		\label{fig:bucket_rbt_cbc}
	\end{subfigure}
	\caption{Bug-finding bucket charts on RBT \\
		\explaincolor{chart_quickcheck}{QuickCheck},
		\explaincolor{chart_hedgehog}{Hedgehog},
		\explaincolor{chart_falsify}{Falsify}}
	\label{fig:bucket_rbt}
\end{figure}

The RBT results in Figure~\ref{fig:bucket_rbt} show a sharper separation between generator families than BST. Naive
type-based generation struggles to produce valid inputs: all three libraries fail to solve half of the RBT tasks.
API-based generation substantially improves
this picture, with QuickCheck solving all tasks and placing them within one second, while
Hedgehog and Falsify miss only a few tasks. Correct-by-construction generation is
in the middle: it improves over type-based generation, but still leaves more tasks unfound
than API-based, especially for Hedgehog and Falsify.

The paired statistical tests agree with the bucket-chart ordering for bug-finding time.
The type-based, API-based, and correct-by-construction comparisons are all significant ($p < 0.001$). QuickCheck is
significantly faster than both Hedgehog and Falsify in all three generator families.
Falsify is also significantly faster than Hedgehog for type-based and correct-by-construction
generation, while the Hedgehog and Falsify comparison for API-based is not significant after Holm
correction.

\begin{wrapfigure}{r}{0.60\textwidth}
	\centering
        \vspace*{-1em}        
	\begin{subfigure}[t]{.48\linewidth}
		\centering
		\includegraphics[width=\linewidth]{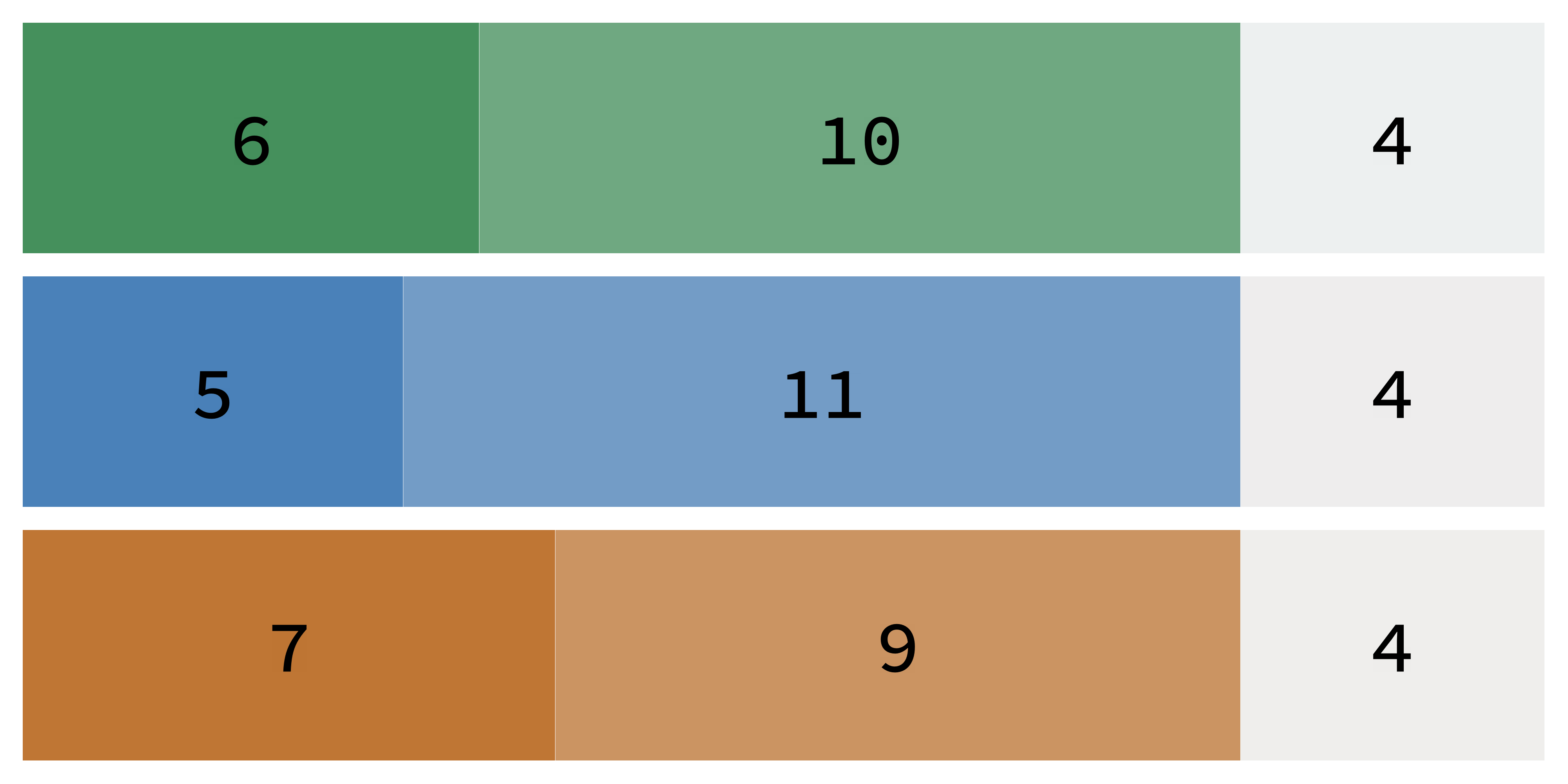}
		\caption{Type-based generation}
		\label{fig:bucket_stlc_vanilla}
	\end{subfigure}\hfill
	\begin{subfigure}[t]{.48\linewidth}
		\centering
		\includegraphics[width=\linewidth]{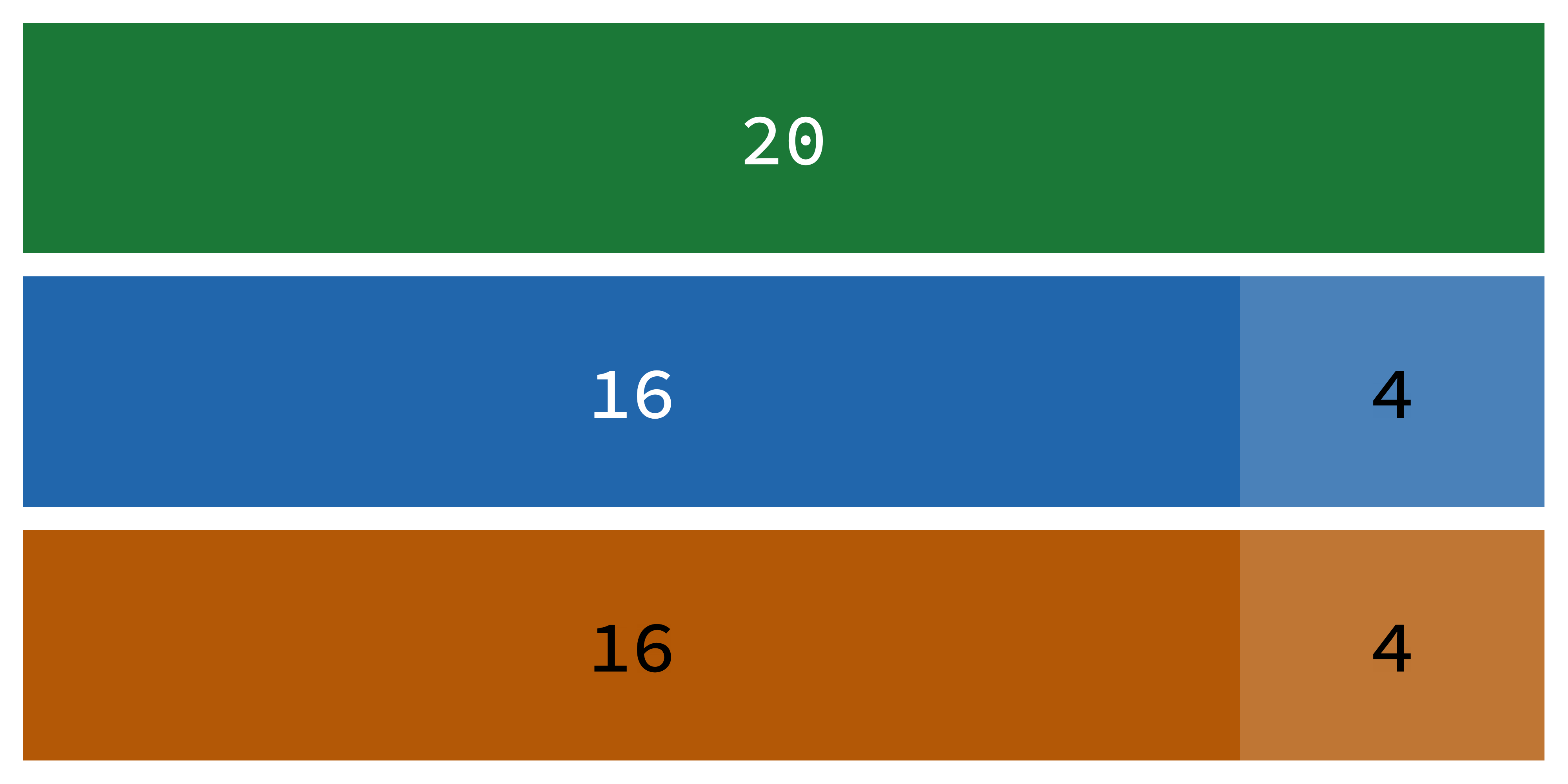}
		\caption{Correct-by-construction}
		\label{fig:bucket_stlc_cbc}
	\end{subfigure}
	\caption{Bug-finding bucket charts on STLC \\
		\explaincolor{chart_quickcheck}{QuickCheck},
		\explaincolor{chart_hedgehog}{Hedgehog},
		\explaincolor{chart_falsify}{Falsify}}
	\label{fig:bucket_stlc}
        \vspace*{-1em}
\end{wrapfigure}

The STLC results in Figure~\ref{fig:bucket_stlc} follow the type-based/correct-by-construction pattern in BST/RBT:
the bucket charts show slower and less complete bug-finding for type-based generation than for correct-by-construction generation.
For type-based generation, the Friedman test
does not find a significant difference in bug-finding time, matching the similar bucket
profiles for QuickCheck, Hedgehog, and Falsify. For correct-by-construction generation,
the difference is significant ($p < 0.001$): QuickCheck is significantly faster than both
Hedgehog and Falsify, while Hedgehog and Falsify are statistically
indistinguishable from each other after Holm correction.

\begin{wrapfigure}{r}{0.60\textwidth}
	\centering
        \vspace*{-1em}
	\begin{subfigure}[t]{.48\linewidth}
		\centering
		\includegraphics[width=\linewidth]{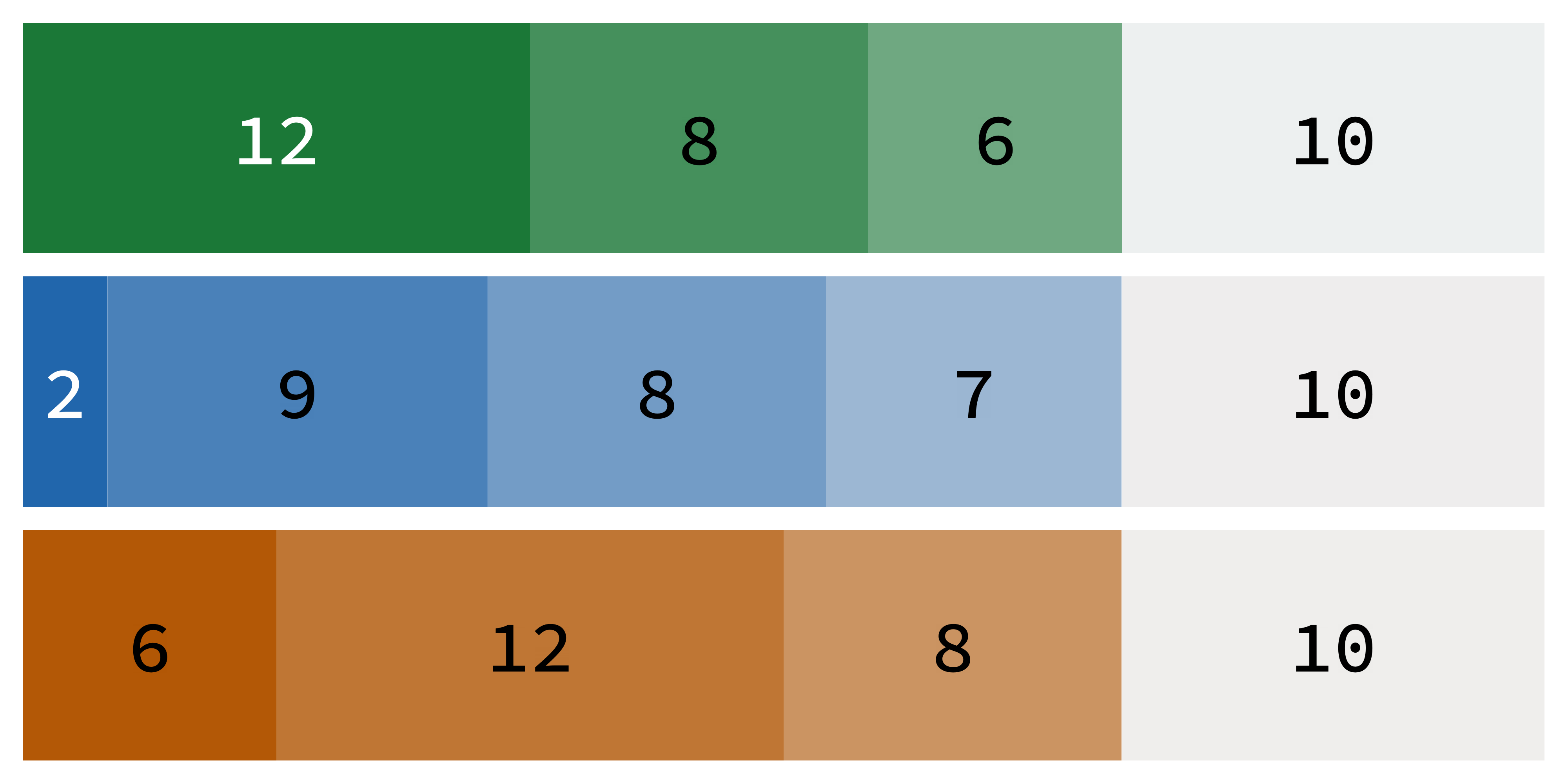}
		\caption{Type-based generation}
		\label{fig:bucket_fsub_vanilla}
	\end{subfigure}\hfill
	\begin{subfigure}[t]{.48\linewidth}
		\centering
		\includegraphics[width=\linewidth]{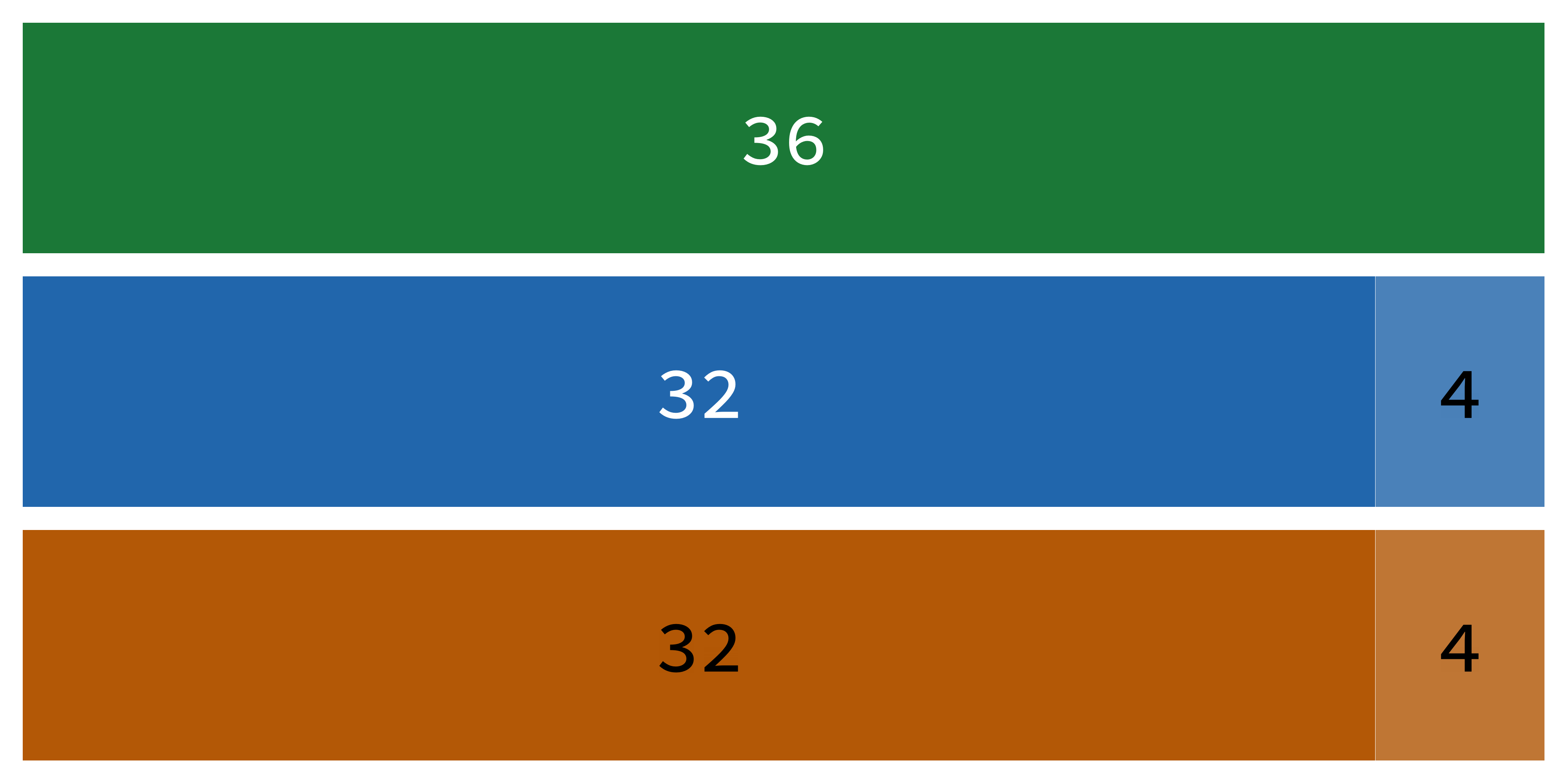}
		\caption{Correct-by-construction}
		\label{fig:bucket_fsub_cbc}
	\end{subfigure}
	\caption{Bug-finding bucket charts on $F_{<:}$ \\
		\explaincolor{chart_quickcheck}{QuickCheck},
		\explaincolor{chart_hedgehog}{Hedgehog},
		\explaincolor{chart_falsify}{Falsify}.}
	\label{fig:bucket_fsub}
        \vspace*{-1em}
\end{wrapfigure}

The $F_{<:}$ results in Figure~\ref{fig:bucket_fsub} mirror STLC but with a larger gap among the naive type-based
generators. Among the
successful runs, QuickCheck has the strongest profile, followed by Falsify, while Hedgehog
places substantially more tasks in slower buckets. Correct-by-construction generation
removes the coverage problem entirely: all generators solve every
task within one second, with QuickCheck placing all tasks in the fastest bucket.

The statistical tests agree with this ordering for bug-finding time. In the type-based
comparison, the Friedman test rejects the null hypothesis of equal medians ($p < 0.001$), QuickCheck is faster
than both Hedgehog and Falsify and Falsify faster than Hedgehog. In the
correct-by-construction comparison, QuickCheck is faster than both
Hedgehog and Falsify, and Falsify is faster than Hedgehog; the results of the statistical
tests can be found in Appendix~\ref{app:stats}.

\subsubsection{Comparison of Shrinking Effectiveness}

\begin{figure}[h!]
	\centering
	\begin{subfigure}[t]{.30\linewidth}
		\centering
		\includegraphics[width=\linewidth]{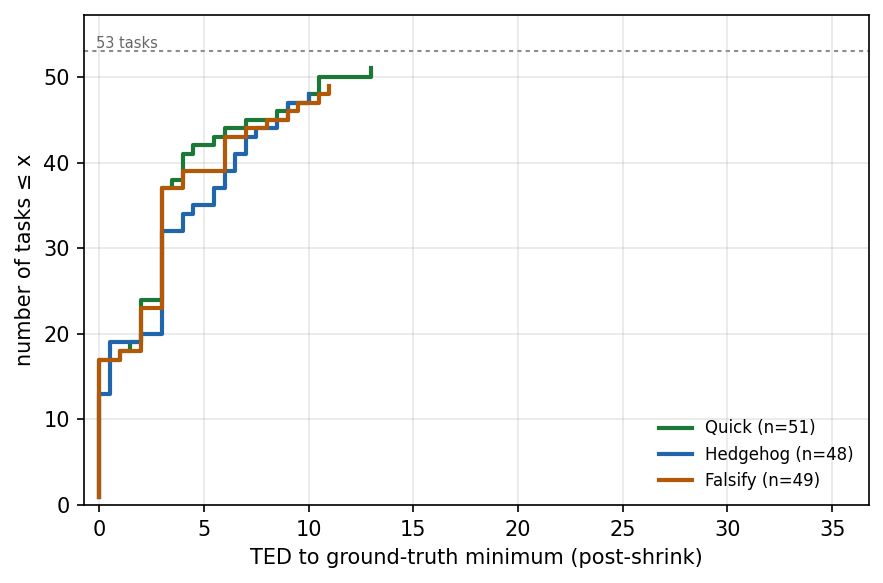}
		\caption{BST, type-based}
	\end{subfigure}
	\begin{subfigure}[t]{.30\linewidth}
		\centering
		\includegraphics[width=\linewidth]{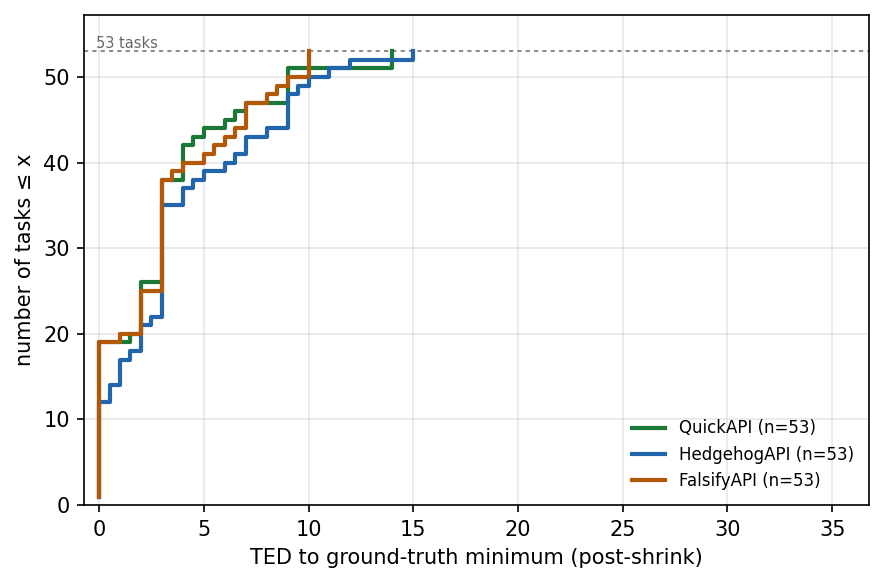}
		\caption{BST, API}
	\end{subfigure}
	\begin{subfigure}[t]{.30\linewidth}
		\centering
		\includegraphics[width=\linewidth]{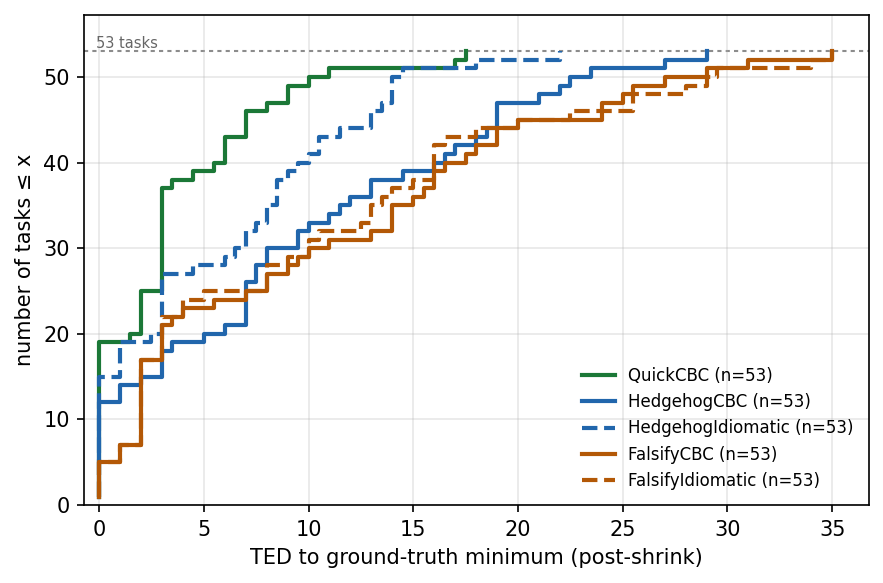}
		\caption{BST, CBC}
	\end{subfigure}\\[0.6em]
	\begin{subfigure}[t]{.30\linewidth}
		\centering
		\includegraphics[width=\linewidth]{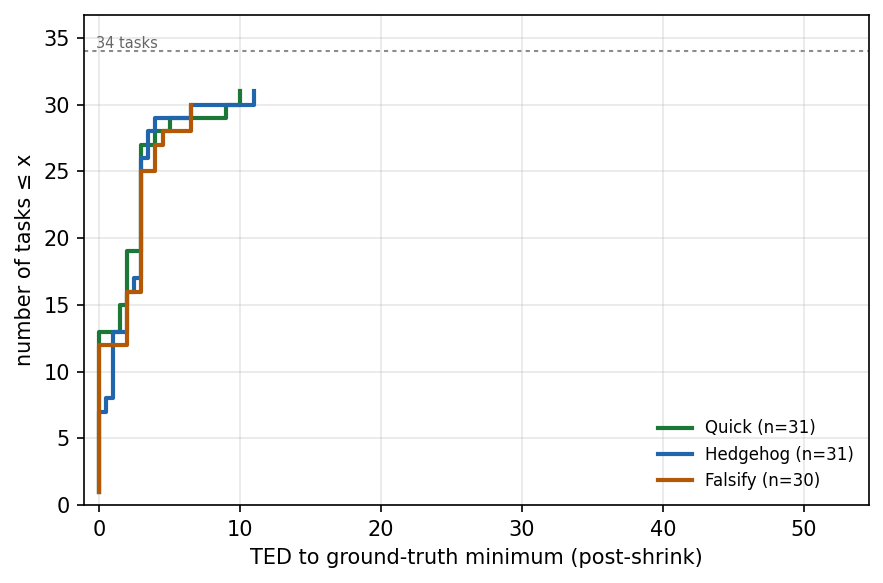}
		\caption{RBT, type-based}
	\end{subfigure}
	\begin{subfigure}[t]{.30\linewidth}
		\centering
		\includegraphics[width=\linewidth]{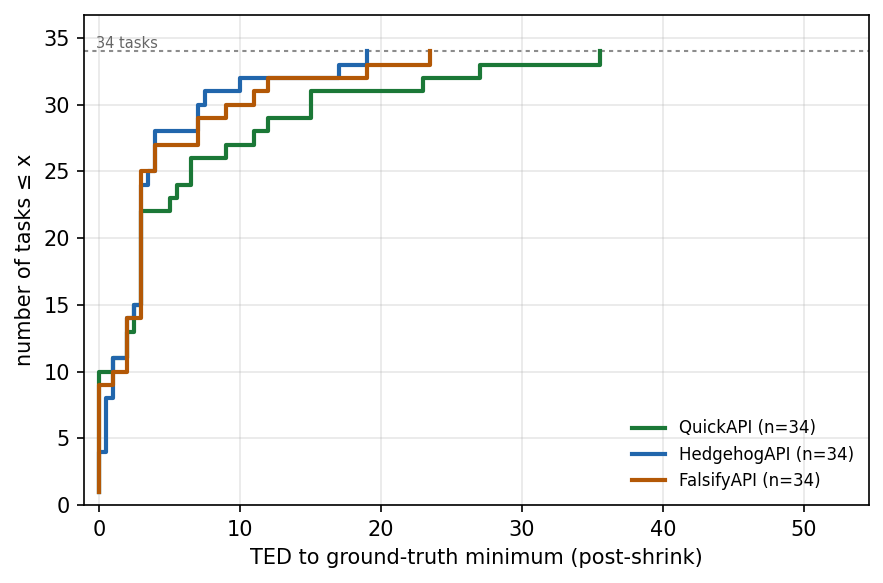}
		\caption{RBT, API}
	\end{subfigure}
	\begin{subfigure}[t]{.30\linewidth}
		\centering
		\includegraphics[width=\linewidth]{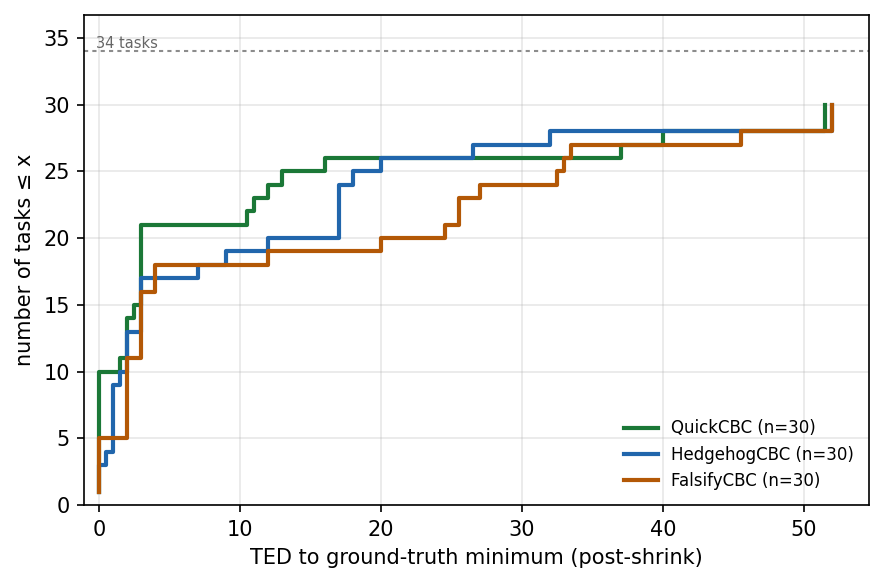}
		\caption{RBT, CBC}
	\end{subfigure}\\[0.6em]
	\begin{subfigure}[t]{.24\linewidth}
		\centering
		\includegraphics[width=\linewidth]{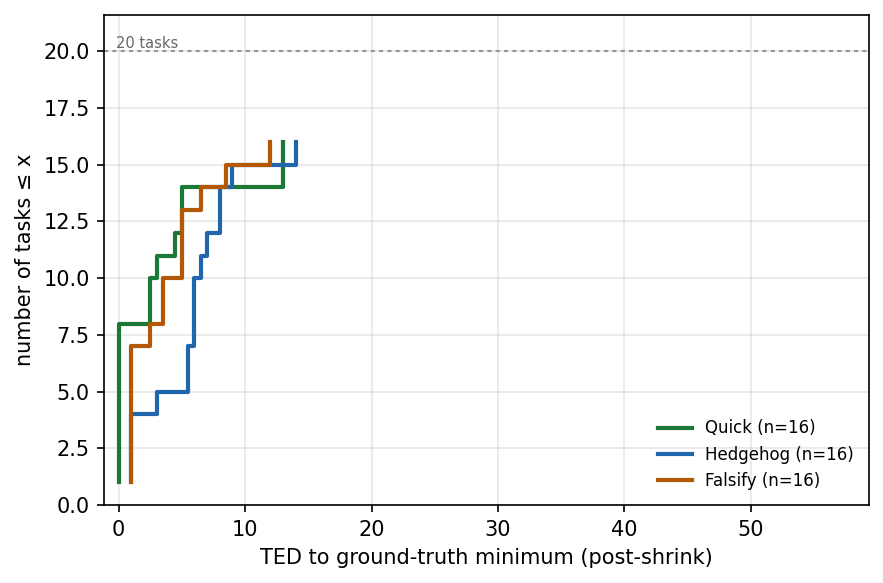}
		\caption{STLC, type-based}
	\end{subfigure}
	\begin{subfigure}[t]{.24\linewidth}
		\centering
		\includegraphics[width=\linewidth]{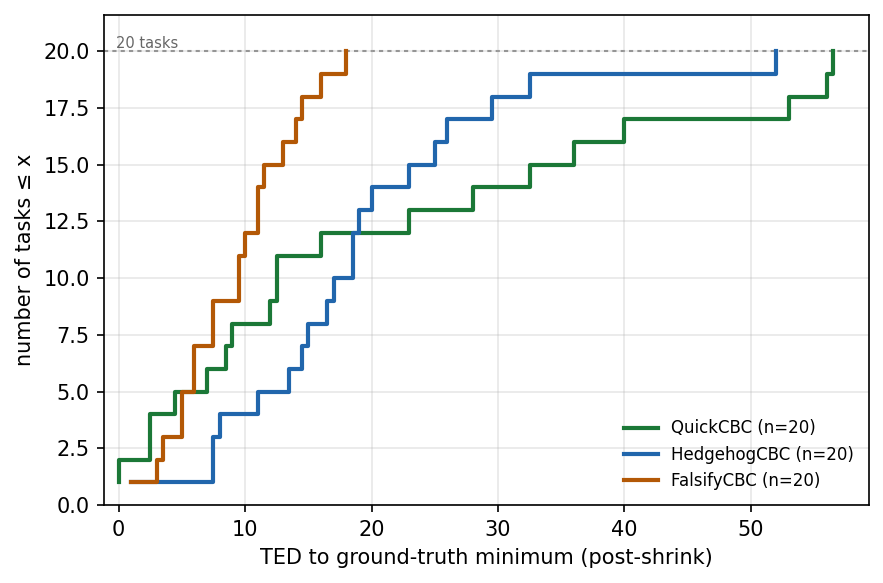}
		\caption{STLC, CBC}
	\end{subfigure}
	\begin{subfigure}[t]{.24\linewidth}
		\centering
		\includegraphics[width=\linewidth]{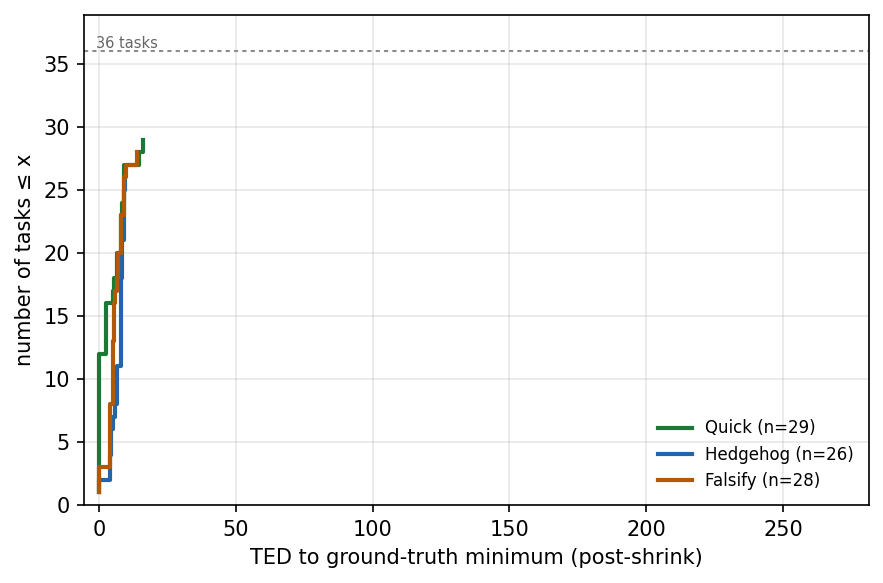}
		\caption{$F_{<:}$, type-based}
	\end{subfigure}
	\begin{subfigure}[t]{.24\linewidth}
		\centering
		\includegraphics[width=\linewidth]{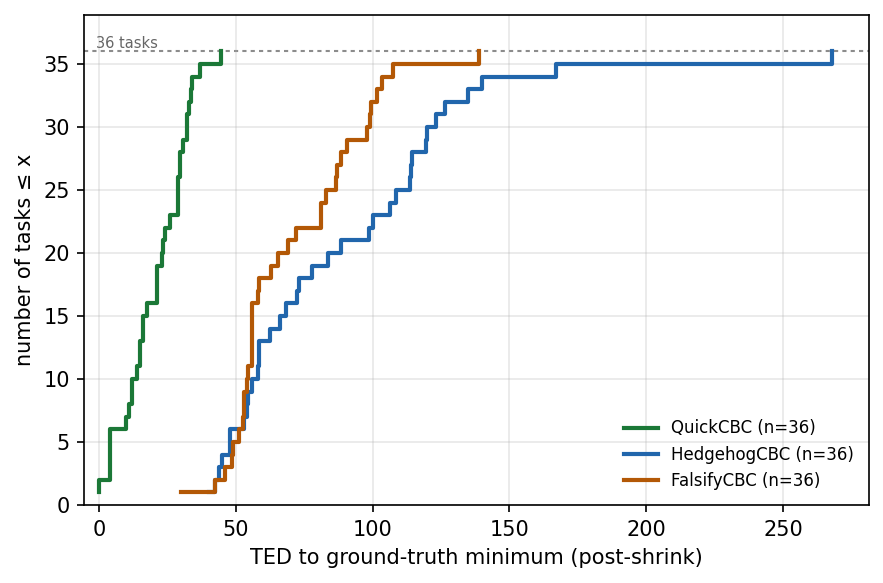}
		\caption{$F_{<:}$, CBC}
	\end{subfigure}
	\caption{CCP of the post-shrink tree edit distance from each strategy's
		minimal counterexample to the ground-truth minimum found by LeanCheck.}
	\label{fig:ecdf_ted_to_gt}
\end{figure}

Figure~\ref{fig:ecdf_ted_to_gt} presents Cumulative Count Plot (CCP) charts denoting shrinking effectiveness
for different libraries where X axis is
the edit distance between the reported shrunk counterexample against the ground truth
minimums computed via exhaustive search using LeanCheck and the Y axis is the number of tasks for the corresponding
workload (top-left is the best, top-right indicates a long tail).

The type-based campaigns show relatively little shrink movement across all four workloads.
All four workloads have preconditions--invariant preservation in BST/RBT and well-typedness
in STLC/$F_{<:}$--that become sparser as input size grows. Consequently, many large generated
inputs are discarded before shrinking begins, and the successful first counterexamples are
already biased toward smaller valid inputs. The measured shrink distance supports this
interpretation: for type-based generators, shrinking reduces the tree edit distance to the
ground-truth minimum by a median of only $3$--$9$ edits per workload (BST $4$, RBT $3$, STLC
$7$, $F_{<:}$ $9$), substantially below the API-based ($41$--$42$) and
correct-by-construction ($18$--$98$) families.

The correct-by-construction and API-based campaigns are more workload-dependent.
On correct-by-construction generation for BST and RBT, QuickCheck's structural shrinker reports
counterexamples closer to the LeanCheck minimum than the integrated shrinkers, while the RBT
API-based comparison is statistically indistinguishable. On STLC, Falsify
reports smaller counterexamples than both QuickCheck and Hedgehog. On $F_{<:}$, QuickCheck
reports the closest counterexamples, followed by Falsify and then Hedgehog.
These results do not support a blanket claim that either structural or integrated shrinking
is always more effective; the generator family and workload both matter.
Due to space constraints, we provide the detailed statistical comparisons in
Appendix~\ref{app:stats}, including the full per-family, per-metric Friedman and post-hoc
Wilcoxon tables.\footnote{For RBT, LeanCheck found ground-truth minima for only 34 tasks in
reasonable time. We exclude the remaining 24 tasks from distance-to-ground-truth comparisons
and leave ground-truth-free shrinking evaluation as future work.}

\begin{figure}[h!]
	\centering
	\begin{subfigure}[t]{.30\linewidth}
		\centering
		\includegraphics[width=\linewidth]{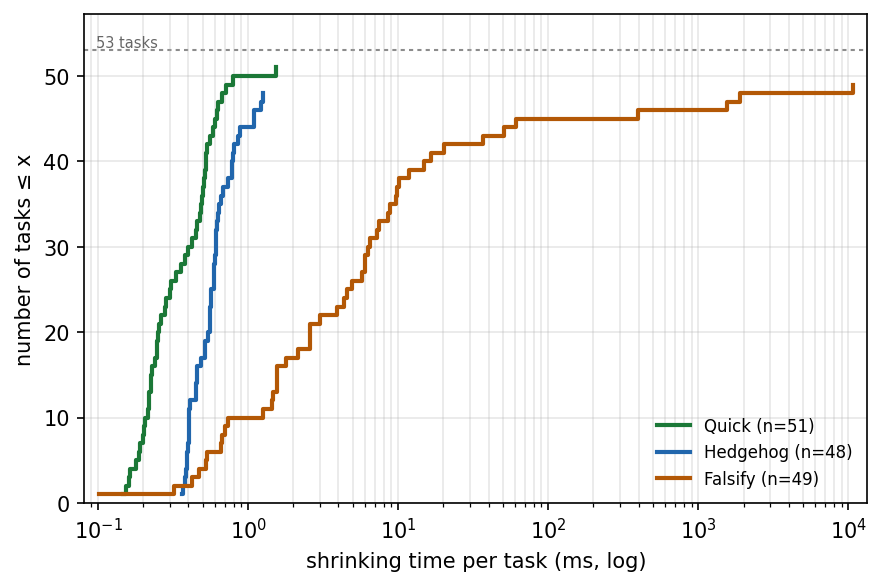}
		\caption{BST, type-based}
	\end{subfigure}
	\begin{subfigure}[t]{.30\linewidth}
		\centering
		\includegraphics[width=\linewidth]{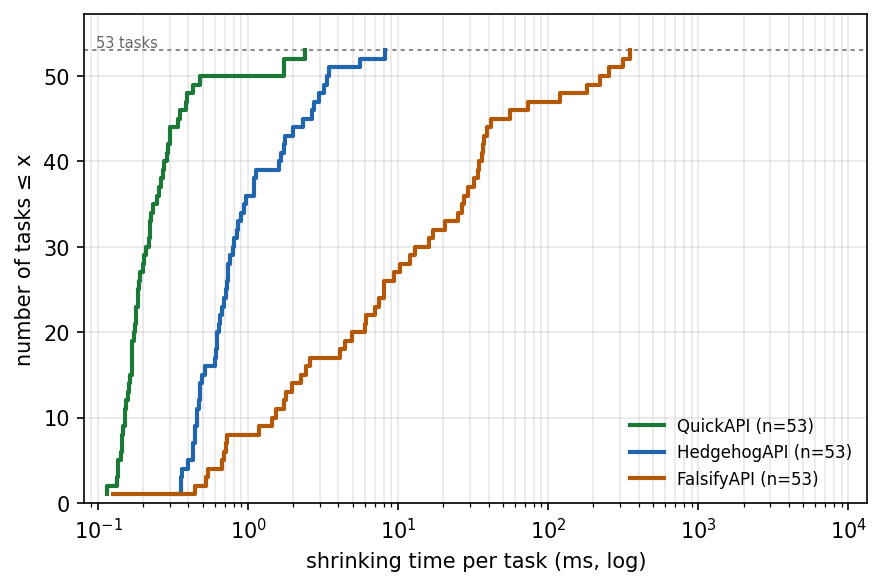}
		\caption{BST, API}
	\end{subfigure}
	\begin{subfigure}[t]{.30\linewidth}
		\centering
		\includegraphics[width=\linewidth]{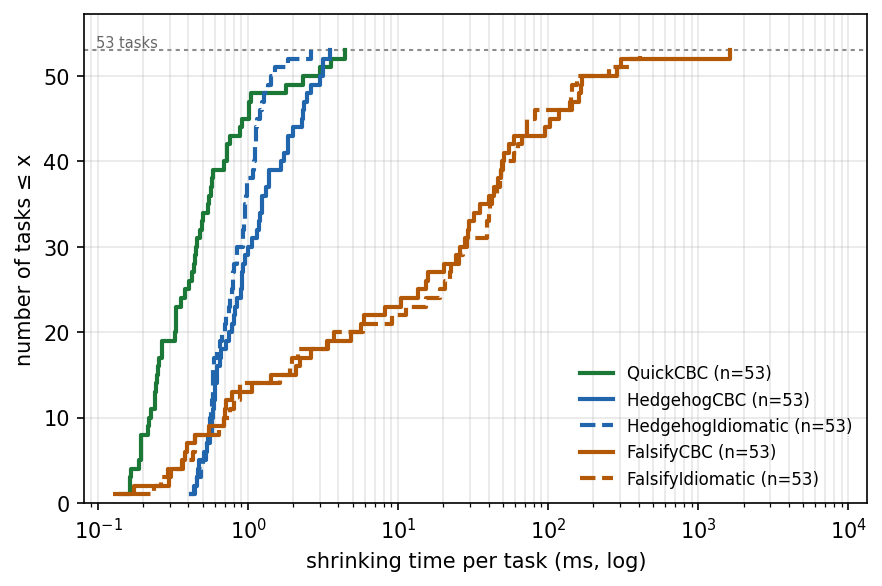}
		\caption{BST, CBC}
	\end{subfigure}\\[0.6em]
	\begin{subfigure}[t]{.30\linewidth}
		\centering
		\includegraphics[width=\linewidth]{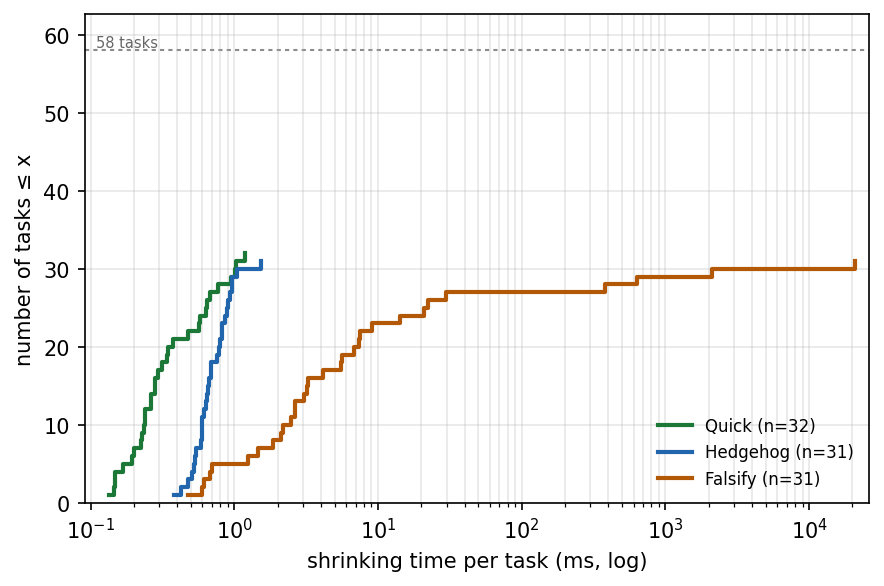}
		\caption{RBT, type-based}
	\end{subfigure}
	\begin{subfigure}[t]{.30\linewidth}
		\centering
		\includegraphics[width=\linewidth]{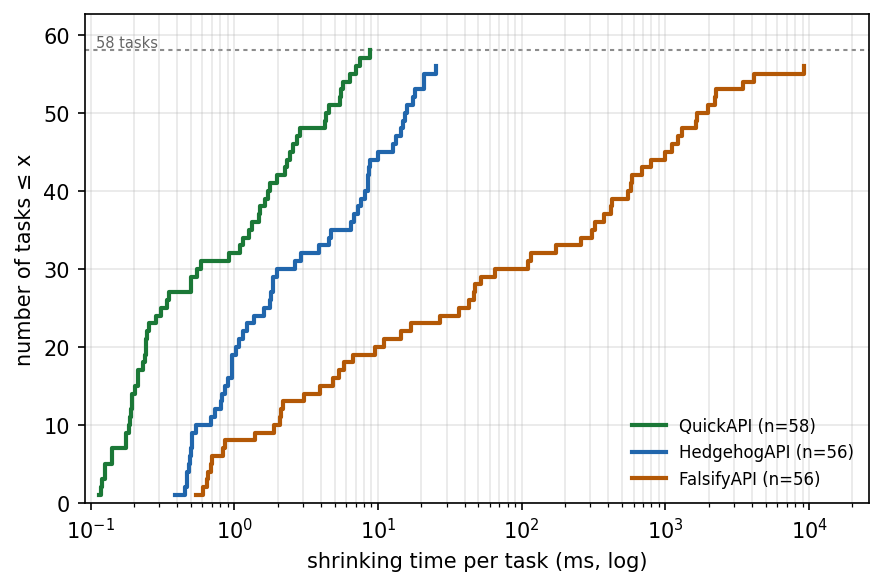}
		\caption{RBT, API}
	\end{subfigure}
	\begin{subfigure}[t]{.30\linewidth}
		\centering
		\includegraphics[width=\linewidth]{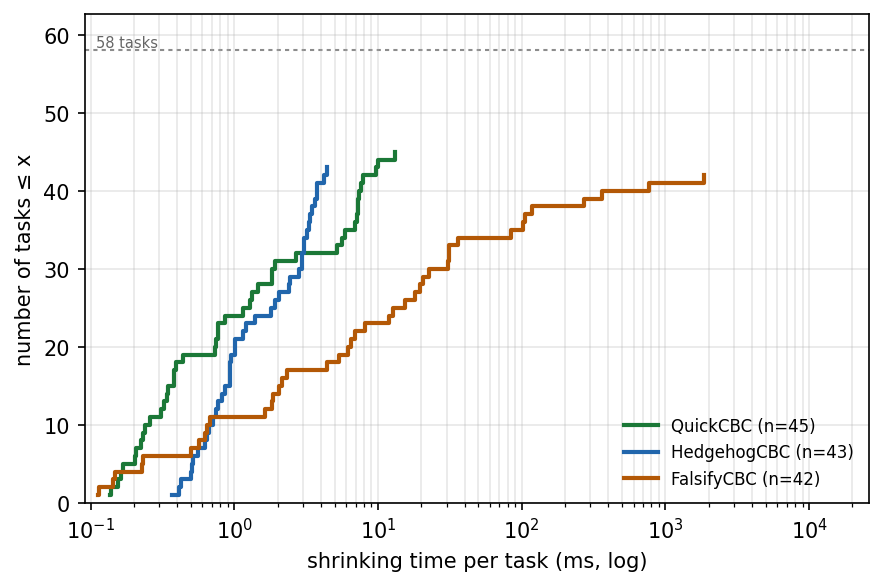}
		\caption{RBT, CBC}
	\end{subfigure}\\[0.6em]
	\begin{subfigure}[t]{.24\linewidth}
		\centering
		\includegraphics[width=\linewidth]{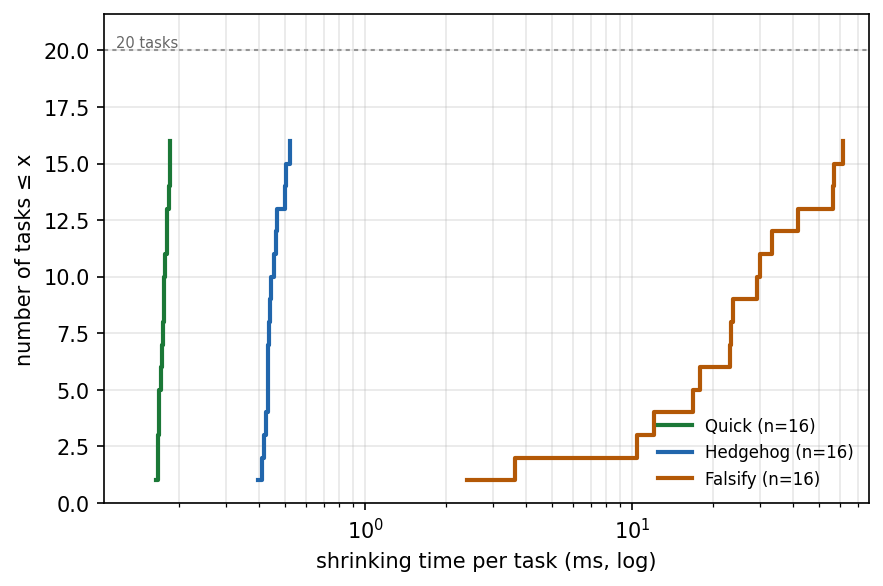}
		\caption{STLC, type-based}
	\end{subfigure}
	\begin{subfigure}[t]{.24\linewidth}
		\centering
		\includegraphics[width=\linewidth]{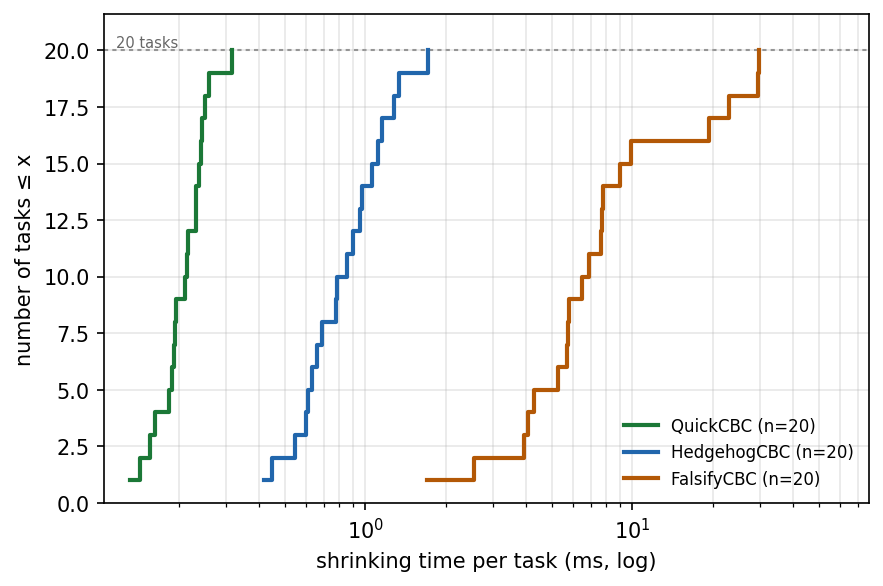}
		\caption{STLC, CBC}
	\end{subfigure}
	\begin{subfigure}[t]{.24\linewidth}
		\centering
		\includegraphics[width=\linewidth]{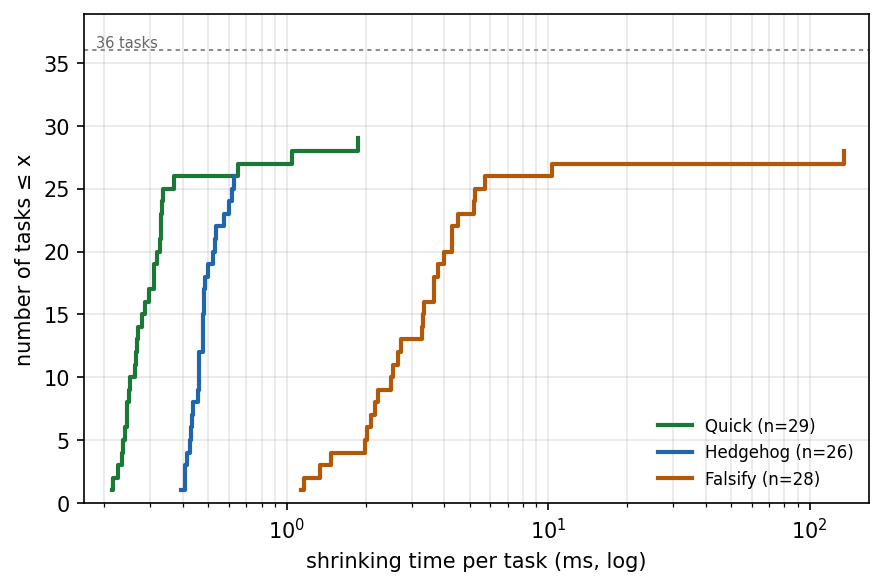}
		\caption{$F_{<:}$, type-based}
	\end{subfigure}
	\begin{subfigure}[t]{.24\linewidth}
		\centering
		\includegraphics[width=\linewidth]{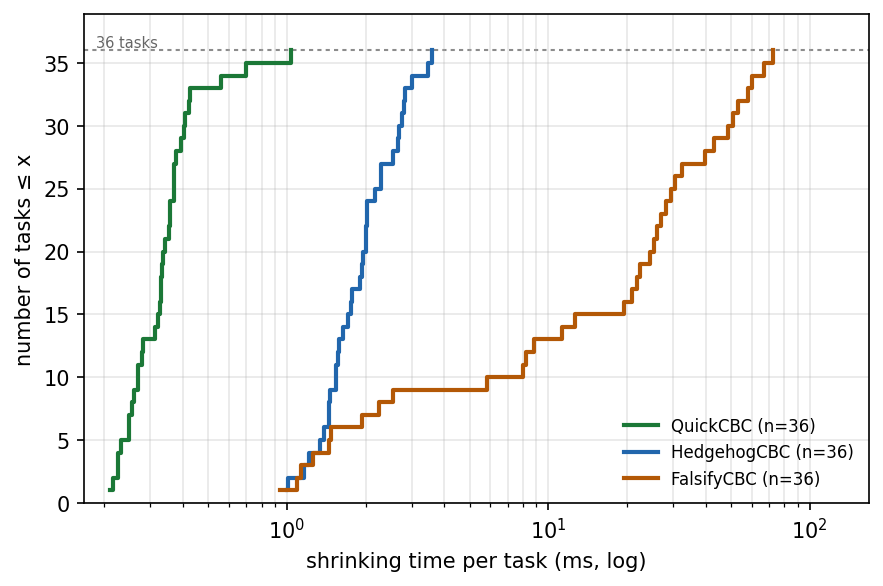}
		\caption{$F_{<:}$, CBC}
	\end{subfigure}
	\caption{CCP of per-task shrinking wall-clock time (ms, log scale)}
	\label{fig:ecdf_time_shrinking}
\end{figure}
\begin{figure}[h!]
	\centering
	\begin{subfigure}[t]{.30\linewidth}
		\centering
		\includegraphics[width=\linewidth]{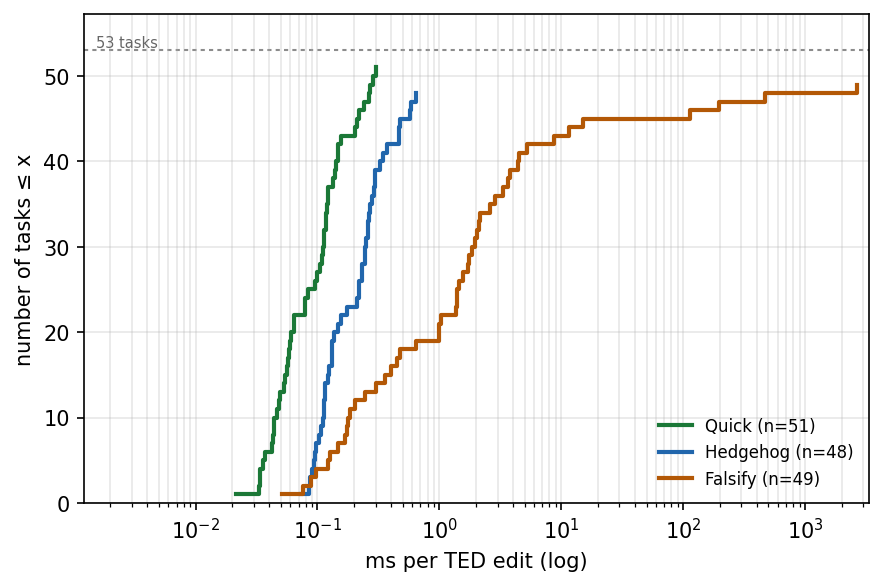}
		\caption{BST, type-based}
	\end{subfigure}
	\begin{subfigure}[t]{.30\linewidth}
		\centering
		\includegraphics[width=\linewidth]{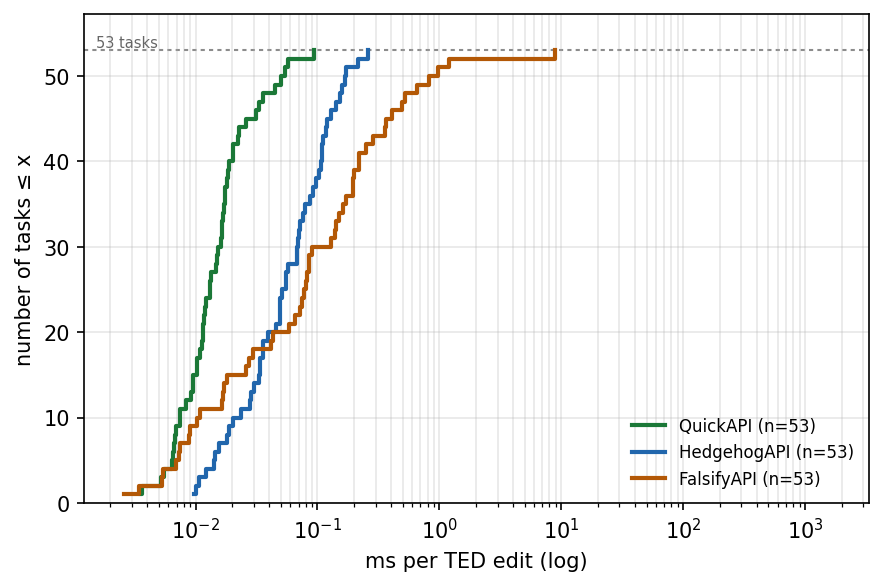}
		\caption{BST, API}
	\end{subfigure}
	\begin{subfigure}[t]{.30\linewidth}
		\centering
		\includegraphics[width=\linewidth]{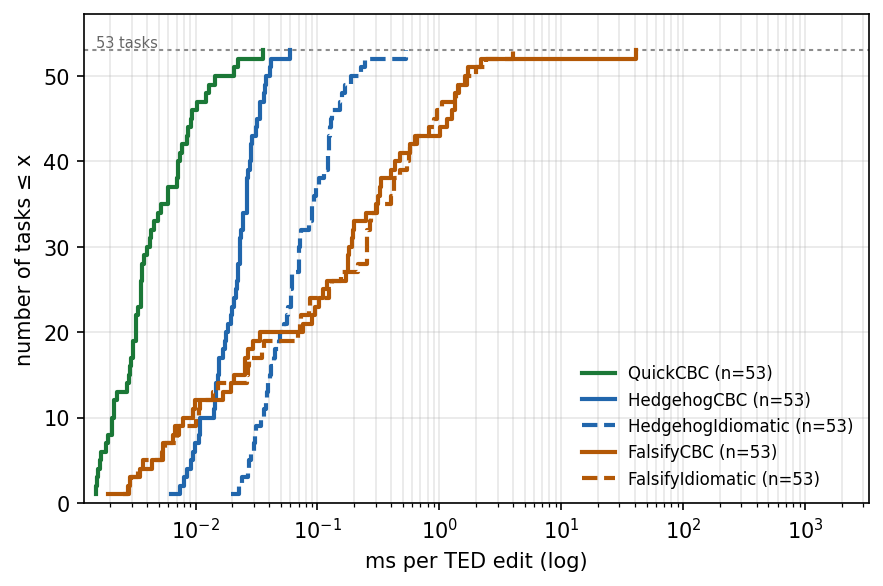}
		\caption{BST, CBC}
	\end{subfigure}\\[0.6em]
	\begin{subfigure}[t]{.30\linewidth}
		\centering
		\includegraphics[width=\linewidth]{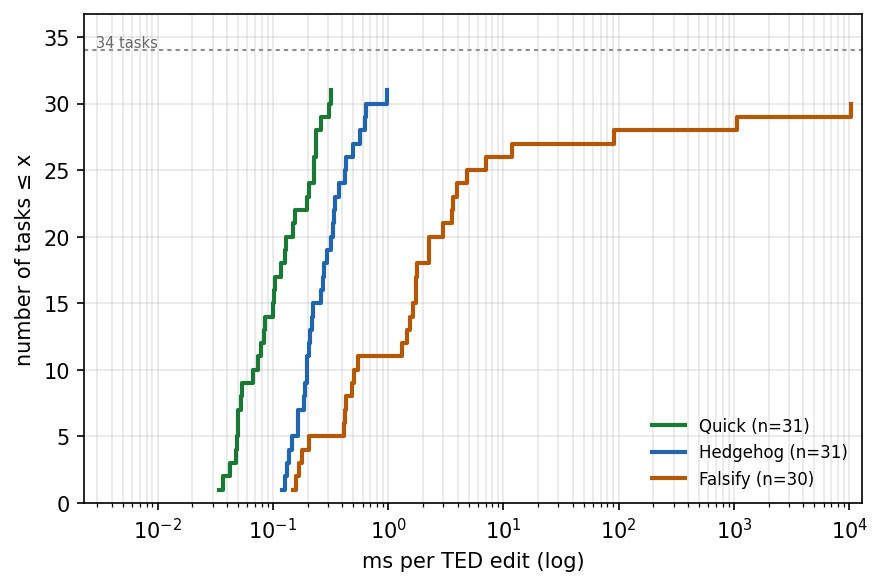}
		\caption{RBT, type-based}
	\end{subfigure}
	\begin{subfigure}[t]{.30\linewidth}
		\centering
		\includegraphics[width=\linewidth]{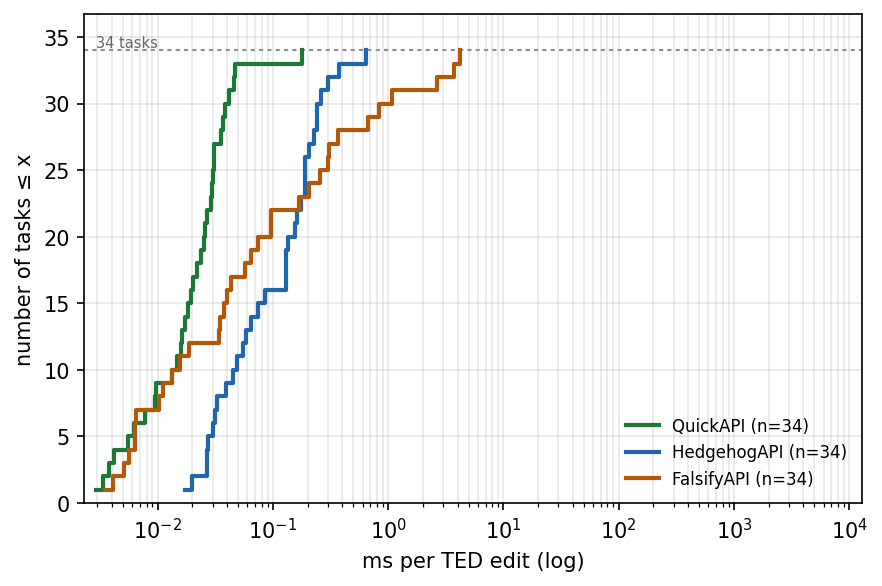}
		\caption{RBT, API}
	\end{subfigure}
	\begin{subfigure}[t]{.30\linewidth}
		\centering
		\includegraphics[width=\linewidth]{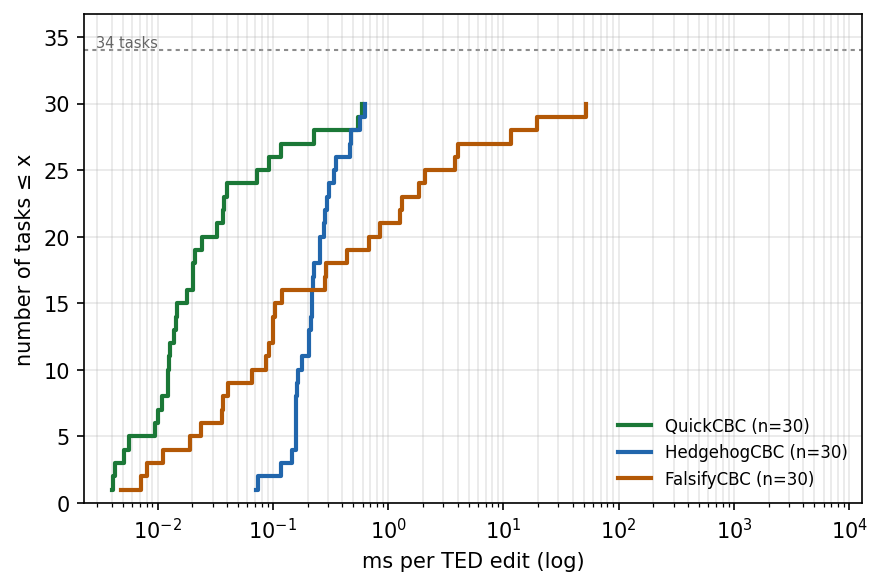}
		\caption{RBT, CBC}
	\end{subfigure}\\[0.6em]
	\begin{subfigure}[t]{.24\linewidth}
		\centering
		\includegraphics[width=\linewidth]{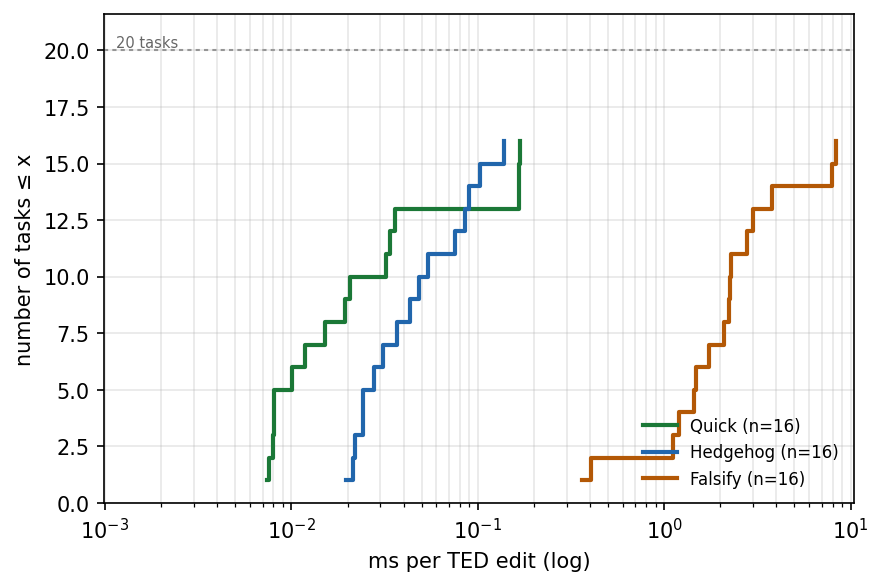}
		\caption{STLC, type-based}
	\end{subfigure}
	\begin{subfigure}[t]{.24\linewidth}
		\centering
		\includegraphics[width=\linewidth]{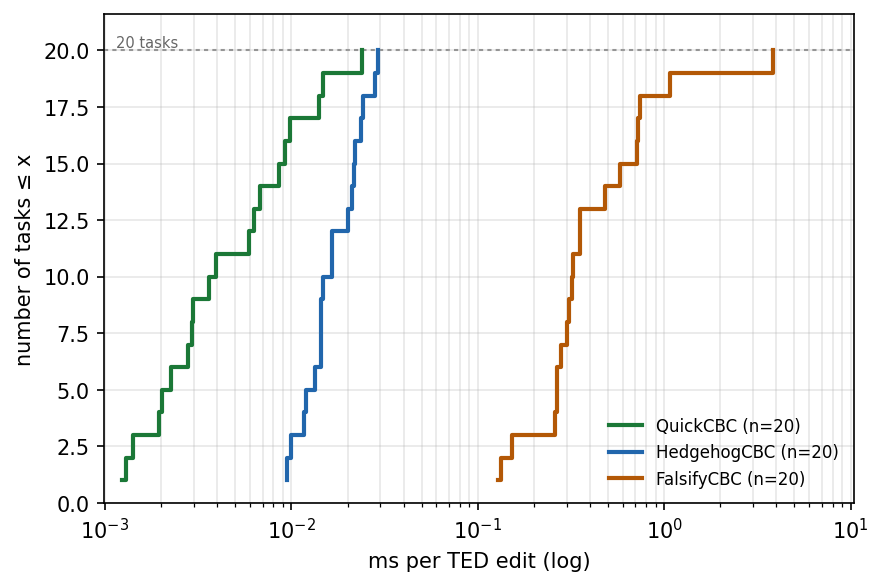}
		\caption{STLC, CBC}
	\end{subfigure}
	\begin{subfigure}[t]{.24\linewidth}
		\centering
		\includegraphics[width=\linewidth]{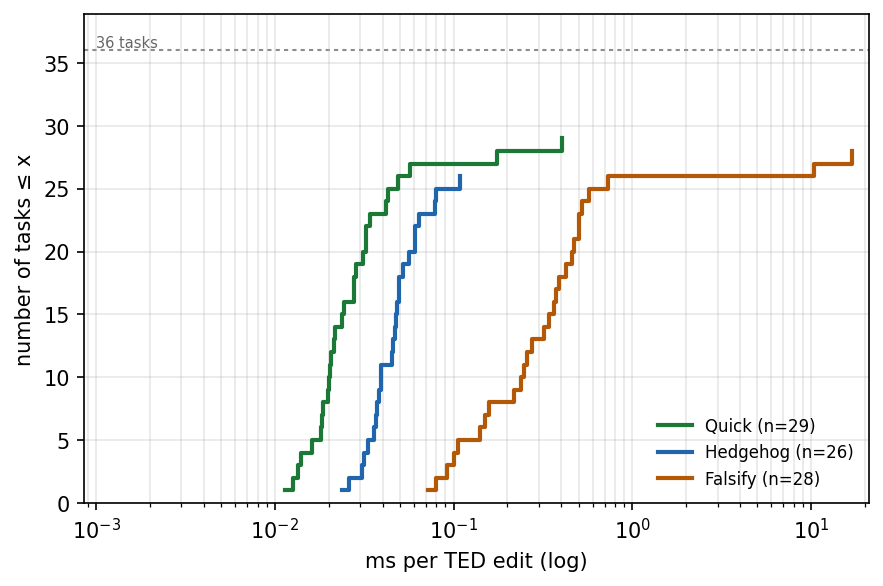}
		\caption{$F_{<:}$, type-based}
	\end{subfigure}
	\begin{subfigure}[t]{.24\linewidth}
		\centering
		\includegraphics[width=\linewidth]{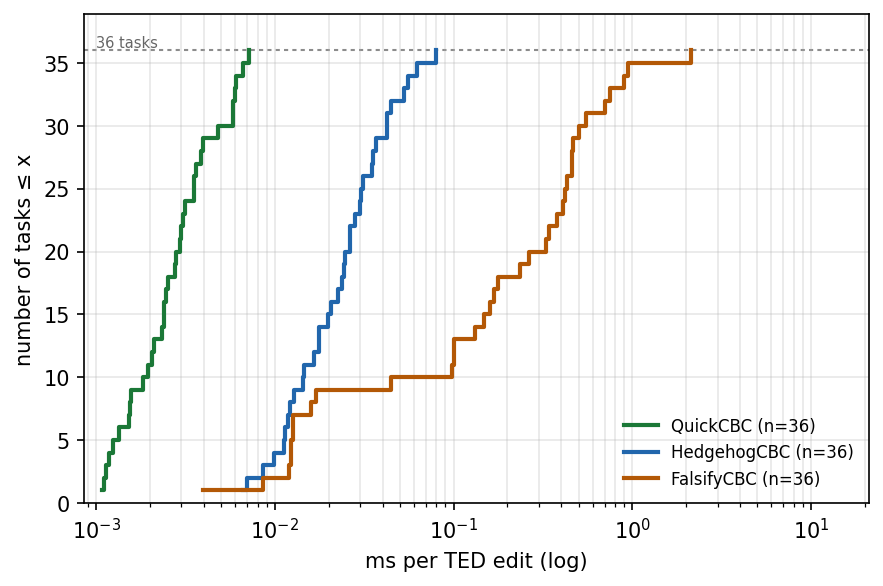}
		\caption{$F_{<:}$, CBC}
	\end{subfigure}
	\caption{CCP of shrinking time per unit of progress: milliseconds spent
		shrinking divided by the tree edit distance reduced (log scale).}
	\label{fig:ecdf_ms_per_edit}
\end{figure}

We follow up with another performance measurement, this time on shrinking itself.
We show both absolute shrinking time in Figure~\ref{fig:ecdf_time_shrinking} as well as time per
edit distance between the original counterexample and the reported shrinking result in Figure~\ref{fig:ecdf_ms_per_edit}
to normalize over different counterexamples reported by each library.

For type-based generators, QuickCheck and Hedgehog have similar shrinking times across the
four workloads, while Falsify is consistently slower and has a longer tail. For a fraction of
tasks, Falsify's shrinking time is several orders of magnitude larger than the others, whereas
QuickCheck and Hedgehog exhibit smaller variation across tasks. This pattern is consistent
with our \emph{shrinking effort} measurements, i.e., the number of executions used to explore
alternatives during shrinking. The libraries expose knobs for tuning shrinking budgets, but
the parameters are not directly comparable: QuickCheck's budget bounds total executions,
whereas Hedgehog and Falsify bound failing executions.
The graphs use each library's default shrinking budget. We also ran no-shrinking
(\texttt{budget = 0}) and fixed-budget (\texttt{budget = 100}) configurations. The
no-shrinking runs let us check the bug-finding overhead of enabling shrinking, while the
fixed-budget runs were intended to standardize effort. The latter did not achieve comparable
effort across libraries because the budget parameters count different events, so we do not
draw conclusions from those runs here.

For correct-by-construction generators, RBT puts QuickCheck and Hedgehog close to each other
against a slower Falsify. On BST, QuickCheck is significantly faster than both Hedgehog and
Falsify. On STLC and $F_{<:}$, the ordering is clearer: QuickCheck is fastest,
Hedgehog is next, and Falsify is slowest.

In an attempt to normalize over shrinking effort, we have also measured time per edit distance between
the original counterexample against the shrunk counterexample as presented in Figure~\ref{fig:ecdf_ms_per_edit}.
The per-edit results are largely consistent with the absolute comparison for QuickCheck and Hedgehog.
For Falsify, the two diverge sharply in the API-based campaigns: its API-based generators
start from far larger pre-shrink counterexamples (median pre-shrink TED = 150 vs = 13-18 for the others),
so the absolute comparison conflated higher shrinking effort with slower shrinking. Normalizing per edit
collapses Falsify's gap to QuickCheck from 49x/95x (BST/RBT) to 6.2x/2.4x. In the type-based and correct-by-construction
campaigns, where Falsify does not start from larger counterexamples, the per-edit and absolute comparisons agree.

\subsection{Discussion}\label{sec:discussion}

We now discuss the implications of these results in context of real world PBT usage.
Namely, an important shortcoming of the original ETNA experiments is the lack of separation of
bug-finding and shrinking time, specifically because bug-finding is a search for an ``unknown unknown'';
the testing process is searching for a bug that potentially does not exist. Once the testing
uncovers a bug, a structure that is inherently finite, and hence has a large, complex but ultimately
finite amount of search space; the shrinking is searching for a ``known unknown'' with an acceptable
false negative result, a bug that is not truly minimal, but still a useful starting ground for the user.
The implication, if one is to accept this argument, is that it is not
acceptable to worsen bug-finding performance at the expense of better shrinking performance, given that
it is justifiable to spend more time on a found bug than searching for a bug that might not exist.
Our results, combined with these arguments, support a narrower conclusion: on these ETNA
workloads, QuickCheck-style structural shrinking remains highly competitive, and often
preferable, on the measured axes of bug-finding time, shrinking time, and distance to known
minima. This should not be read as a general recommendation against integrated shrinking.
Hedgehog and Falsify reduce the need to write separate shrinkers and can preserve generator
invariants by construction; those usability benefits are real, but they are not directly
measured by our experiments.

This result must also be taken in context; the representativeness of ETNA workloads in terms of
precondition sparsity can significantly affect structural shrinking.
Moreover, a large chunk of the experiment results in this section depend on the execution time. ETNA-Haskell
workloads are small programs; the large search space for both generation and shrinking indicates their quality
in comparing effectiveness of the algorithms for generation and shrinking, but any indications of performance are
potentially biased by the domination of the time spent in the PBT library as opposed to executing the tests themselves.
As execution times get longer for workloads, the PBT library performance is less critical in the overall execution, whereas
the quality of the algorithms will be more important.

%

An important metric for this discussion is \emph{sample-efficiency}: how many candidates a
shrinker must test during the shrinking phase, relative to the progress it makes. A
sample-efficient shrinker should scale better as property execution time grows, because it
spends fewer executions on candidates that do not reproduce the failure. One argument for
integrated shrinking is that it should improve sample-efficiency for properties with
preconditions, because generator invariants are preserved by construction. Our measurements
partly support this argument for Hedgehog: for correct-by-construction generators, Hedgehog
preserves a 26--56\% failure rate across shrinking steps, compared to 5--12\% for QuickCheck
and roughly 2--3\% for Falsify. The stark contrast between Hedgehog and Falsify is important:
integrated shrinking can avoid invalid candidates introduced by structural shrinking, but integrated shrinking alone does not
guarantee high sample-efficiency. Moreover, it is only one part of the
trade-off; in these workloads, Hedgehog's higher failure rate during shrinking does not
translate into better shrinking time or consistently better final counterexamples.


\subsection{Limitations and Threats to Validity}\label{threats}

As we worked on our measurements, we have seen that presenting concrete measurements that provide insights
is a genuinely complex task. The generators we use are based on the original QuickCheck generators written
for ETNA, and are therefore likely to favor QuickCheck. We attempted to resolve this possibility by going
through Hedgehog and Falsify repositories to find idiomatic alternatives, which resulted in the idiomatic CBC generators
for BST that we report on.

Measuring performance in Haskell requires forcing strictness in the runners, which might
result in scenarios where our measurements do not completely agree with empirical real world usage without
the forced strictness. We will release the whole experimentation suite with the deanonymized submission
as a reproducible benchmark in order to allow for outside contribution to correct any measurement errors that might
have been the result of an oversight on the measurements. It is also very hard to control for effort across libraries,
we simply could not set constant shrinking effort across libraries, and therefore opted for the default configurations
for our final experiments.

Lastly, tree edit distance to ground truth minimums can be a noisy metric. For instance, it treats symmetric counterexamples
(A-B vs B-A) differently, when they might in fact be equivalent. The minimum also relies on the exhaustive search algorithm,
a different enumeration strategy might (and does) reach different inputs. As such, we also report on counterexample sizes
in the statistical tests in Appendix~\ref{app:stats} which we observed resulted in broadly similar results to the edit distance
experiments. Additionally, one might argue that the edit distance metric may be closer to the structural notion of size exposed by
QuickCheck-style shrinkers than to the randomness-buffer or choice-tree search spaces used by integrated shrinkers. The ultimate
measure of shrinking effectiveness is debugging time, which would require a comprehensive user study that we deemed
way out of scope for this paper.

\section{Related Work}
\label{sec:related}

The closest related work on PBT shrinking is the study of the Hypothesis
reducer~\cite{HypothesisShrinking}. That work evaluates test-case reduction using case
studies, mean input size, and the number of executions in the shrinking step. We build on
that perspective by measuring distance to known minima, separating bug-finding time from
shrinking time, and comparing several Haskell libraries and generator families on the same
ETNA workloads.

Shrinking is part of a broader family of techniques that iteratively simplify an input or
program while preserving a property of interest. In \emph{test-case reduction}, the preserved
property is a triggering failure: delta debugging introduced iterative simplification of
failing inputs~\cite{zeller2002simplifying}, followed by optimized techniques leveraging
the hierarchical structure of the test case~\cite{misherghi2006hdd}, the formal syntax~\cite{sun2018perses},
or the tree structure~\cite{herfert2017automatically}. Researchers have also proposed
probabilistic~\cite{wang2021probabilistic} and weighted~\cite{zhou2025wdd} alternatives
to delta-debugging, as well as domain-specific reducers for compiler
testing~\cite{RegehrCCEEY12,donaldson2021test}, SQL queries~\cite{lin2024sqless} and
dependency graphs~\cite{kalhauge2019binary}. Related lines simplify programs under different objectives:
\emph{program trimming} preserves equi-safety to scale static analyzers~\cite{ferles2017failure},
while \emph{program and container debloating} preserves intended functionality to reduce attack
surface~\cite{rastogi2017cimplifier,heo2018effective}. Across these settings, evaluation typically
reports size reduction and reduction time on benchmark suites of recorded failures or programs.


\section{Conclusion and Future Work}
\label{sec:conc}

We presented an experience report on evaluating shrinking across QuickCheck, Hedgehog, and
Falsify on four ETNA workloads. The main empirical lesson is that integrated shrinking does
not automatically dominate structural shrinking. In these experiments, QuickCheck-style
structural shrinking is usually faster and often competitive in final counterexample quality,
while integrated shrinking shows workload- and generator-dependent behavior. The main
methodological lesson is that shrinking should be evaluated separately from bug-finding:
final counterexample quality, shrink time, per-edit cost, sample-efficiency, and generator
choice expose different trade-offs.

Future work should broaden the benchmark suite beyond ETNA's Haskell
workloads, and perhaps investigate alternative algorithms for shrinking.
In this paper, we focused on evaluating the out-of-the-box shrinking behavior;
each system allows for different degrees of customizing shrinking, and evaluating
the effect (and effort) of those is particularly interesting future work.
Finally, a user study measuring debugging time would also help
connect structural metrics, such as tree edit distance, shrinking
ratio or counterexample size to the outcome developers ultimately care
about.

  \clearpage
\appendix

\section{Binary-Search Tree Generators} \label{app:bst-generators}

\begin{hask}
gen_bst_falsify :: Int -> Int -> Int -> Gen BST
gen_bst_falsify depth lo hi
  | depth <= 0 || lo + 1 >= hi = pure E
  | otherwise =
      Gen.frequency
        [ (1, pure E)
        , ( 3
          , do
              k <- Gen.int (Range.withOrigin (lo + 1, hi - 1) 0)
              v <- Gen.int (Range.withOrigin (-1000, 1000) 0)
              left <- gen_bst_falsify (depth - 1) lo k
              right <- gen_bst_falsify (depth - 1) k hi
              pure (T left (Key k) (Val v) right)
          )
        ]
\end{hask}

\begin{hask}
gen_bst_hedgehog :: Int -> Int -> Int -> HH.Gen BST
gen_bst_hedgehog depth lo hi
  | depth <= 0 || lo + 1 >= hi = pure E
  | otherwise =
      Gen.frequency
        [ (1, pure E)
        , ( 3
          , do
              k <- Gen.int (Range.linearFrom 0 (lo + 1) (hi - 1))
              v <- Gen.int (Range.linearFrom 0 (-1000) 1000)
              left <- gen_bst_hedgehog (depth - 1) lo k
              right <- gen_bst_hedgehog (depth - 1) k hi
              pure (T left (Key k) (Val v) right)
          )
        ]
\end{hask}

\begin{hask}
gen_bst_quickcheck :: Int -> Int -> Int -> Gen BST
gen_bst_quickcheck depth lo hi
  | depth <= 0 || lo + 1 >= hi = pure E
  | otherwise =
      frequency
        [ (1, pure E)
        , ( 3
          , do
              k <- chooseInt (lo + 1, hi - 1)
              v <- chooseInt (-1000, 1000)
              left <- gen_bst_quickcheck (depth - 1) lo k
              right <- gen_bst_quickcheck (depth - 1) k hi
              pure (T left (Key k) (Val v) right)
          )
        ]
\end{hask}

The two \emph{idiomatic} variants use each framework's native recursive
combinator instead of a manual depth counter. The Hedgehog variant uses
\texttt{Gen.recursive}\cite{hedgehogRecursive}, which gates the recursion on Hedgehog's own
\texttt{Size} parameter:

\begin{hask}
gen_bst_hedgehog_idiomatic :: Int -> Int -> HH.Gen BST
gen_bst_hedgehog_idiomatic lo hi
  | lo + 1 >= hi = pure E
  | otherwise =
      Gen.recursive Gen.choice
        [ pure E ]
        [ do
            k <- Gen.int (Range.linearFrom 0 (lo + 1) (hi - 1))
            v <- Gen.int (Range.linearFrom 0 (-1000) 1000)
            left  <- gen_bst_hedgehog_idiomatic lo k
            right <- gen_bst_hedgehog_idiomatic k hi
            pure (T left (Key k) (Val v) right)
        ]
\end{hask}

The Falsify variant mimics \texttt{Test.Falsify.Generator.bst}\cite{falsifyBSTGenerator}: it keeps
the same generation distribution as \texttt{gen\_bst\_falsify}, but wraps
every recursive subtree with the canonical \texttt{firstThen id (const E)}
subtree-promotion pattern, so the shrinker has an explicit
subtree-collapse candidate at each level.

\begin{hask}
collapseSubtree :: Gen BST -> Gen BST
collapseSubtree g = Gen.firstThen id (const E) <*> g

gen_bst_falsify_idiomatic :: Int -> Int -> Int -> Gen BST
gen_bst_falsify_idiomatic depth lo hi
  | depth <= 0 || lo + 1 >= hi = pure E
  | otherwise =
      Gen.frequency
        [ (1, pure E)
        , ( 3
          , do
              k <- Gen.int (Range.between (lo + 1, hi - 1))
              v <- Gen.int (Range.withOrigin (-1000, 1000) 0)
              left  <- collapseSubtree (gen_bst_falsify_idiomatic (depth - 1) lo k)
              right <- collapseSubtree (gen_bst_falsify_idiomatic (depth - 1) k hi)
              pure (T left (Key k) (Val v) right)
          )
        ]
\end{hask}

\section{Ground-Truth Minimal Counterexamples}\label{app:groundtruth}

The tables below list, for every (property, mutation) task of each workload, the minimal counterexample established by the exhaustive deterministic LeanCheck search. These are the references against which the tree-edit-distance metric in Section~\ref{sec:eval} is computed. A dash (---) marks a task for which the deterministic search did not establish a ground truth.

\begin{longtable}{@{}l l >{\ttfamily\footnotesize}p{0.46\linewidth}@{}}
\caption{Ground-truth minimal counterexamples for BST (53 of 53 tasks solved by the deterministic LeanCheck search).}\label{tab:gt-bst}\\
\toprule
\normalfont Property & \normalfont Mutation & \normalfont Minimal counterexample \\
\midrule
\endfirsthead
\multicolumn{3}{@{}l}{\footnotesize\itshape Table~\ref{tab:gt-bst}, BST, continued}\\
\toprule
\normalfont Property & \normalfont Mutation & \normalfont Minimal counterexample \\
\midrule
\endhead
\bottomrule
\endlastfoot
DeleteInsert & insert\_1 & ((T (E) 1 0 (E)),0,0,0) \\
InsertInsert & insert\_1 & ((E),0,1,0,0) \\
InsertModel & insert\_1 & ((T (E) 0 0 (E)),1,0) \\
InsertPost & insert\_1 & ((T (E) 0 0 (E)),1,0,0) \\
InsertUnion & insert\_1 & ((E),(T (E) 0 0 (E)),1,0) \\
UnionDeleteInsert & insert\_1 & ((T (E) 0 0 (E)),(E),1,0) \\
DeleteInsert & insert\_2 & ((T (E) 0 0 (E)),0,1,0) \\
InsertDelete & insert\_2 & ((T (E) 0 0 (E)),1,0,0) \\
InsertInsert & insert\_2 & ((E),0,1,0,0) \\
InsertModel & insert\_2 & ((T (E) 0 0 (E)),1,0) \\
InsertPost & insert\_2 & ((T (E) 0 0 (E)),1,0,1) \\
InsertUnion & insert\_2 & ((E),(T (E) 0 0 (E)),1,0) \\
UnionDeleteInsert & insert\_2 & ((T (E) 0 0 (E)),(E),1,0) \\
InsertDelete & insert\_3 & ((T (E) 0 0 (E)),0,0,1) \\
InsertInsert & insert\_3 & ((E),0,0,0,1) \\
InsertModel & insert\_3 & ((T (E) 0 0 (E)),0,1) \\
InsertPost & insert\_3 & ((T (E) 0 0 (E)),0,0,1) \\
InsertUnion & insert\_3 & ((E),(T (E) 0 0 (E)),0,1) \\
UnionDeleteInsert & insert\_3 & ((T (E) 0 0 (E)),(E),0,1) \\
DeleteDelete & delete\_4 & ((T (T (T (E) -1 0 (E)) 0 0 (E)) 1 0 (E)),0,1) \\
DeleteInsert & delete\_4 & ((E),0,1,0) \\
DeleteModel & delete\_4 & ((T (E) 0 0 (E)),1) \\
DeletePost & delete\_4 & ((T (E) 0 0 (E)),1,0) \\
DeleteUnion & delete\_4 & ((T (E) 0 0 (E)),(T (E) 1 0 (E)),0) \\
InsertDelete & delete\_4 & ((E),0,1,0) \\
UnionDeleteInsert & delete\_4 & ((T (E) 0 0 (E)),(E),1,0) \\
DeleteDelete & delete\_5 & ((T (E) 0 0 (T (E) 1 0 (E))),0,1) \\
DeleteInsert & delete\_5 & ((T (E) 1 0 (E)),0,0,0) \\
DeleteModel & delete\_5 & ((T (T (E) 0 0 (E)) 1 0 (E)),0) \\
DeletePost & delete\_5 & ((T (T (E) 0 0 (E)) 1 0 (E)),0,0) \\
DeleteUnion & delete\_5 & ((T (E) 1 0 (E)),(T (E) 0 0 (E)),0) \\
UnionDeleteInsert & delete\_5 & ((T (T (E) 0 0 (E)) 1 0 (E)),(E),0,1) \\
DeleteUnion & union\_6 & ((T (E) 0 0 (E)),(T (E) 0 0 (E)),0) \\
InsertUnion & union\_6 & ((E),(T (E) 0 0 (E)),0,0) \\
UnionDeleteInsert & union\_6 & ((T (E) 0 0 (E)),(T (E) 0 0 (E)),0,0) \\
UnionModel & union\_6 & ((T (E) 0 0 (E)),(T (E) 0 0 (E))) \\
UnionPost & union\_6 & ((T (E) 1 0 (E)),(T (E) 0 0 (E)),0) \\
UnionUnionAssoc & union\_6 & ((T (E) 0 0 (E)),(T (E) 0 0 (E)),(T (E) 0 0 (E))) \\
UnionUnionIdem & union\_6 & (T (E) 0 0 (E)) \\
UnionValid & union\_6 & ((T (E) 0 0 (E)),(T (E) 0 0 (E))) \\
DeleteUnion & union\_7 & ((T (E) 0 0 (E)),(T (T (E) 0 0 (E)) 1 0 (E)),0) \\
InsertUnion & union\_7 & ((E),(T (T (E) 0 0 (E)) 1 0 (E)),0,0) \\
UnionDeleteInsert & union\_7 & ((T (E) 0 0 (E)),(T (T (E) 0 0 (E)) 1 0 (E)),0,0) \\
UnionModel & union\_7 & ((T (E) 0 0 (E)),(T (T (E) 0 0 (E)) 1 0 (E))) \\
UnionPost & union\_7 & ((T (T (E) 0 0 (E)) 1 0 (E)),(T (E) 0 1 (E)),0) \\
UnionUnionAssoc & union\_7 & ((T (E) 0 0 (E)),(T (E) 0 0 (E)),(T (E) -1 0 (E))) \\
UnionValid & union\_7 & ((T (E) 0 0 (E)),(T (T (E) 0 0 (E)) 1 0 (E))) \\
DeleteUnion & union\_8 & ((T (E) 1 0 (E)),(T (E) 0 0 (T (E) 1 1 (E))),0) \\
InsertUnion & union\_8 & ((T (E) 1 0 (E)),(T (E) 0 0 (E)),0,1) \\
UnionDeleteInsert & union\_8 & ((T (T (E) 0 0 (E)) 1 0 (E)),(T (E) 0 1 (E)),1,0) \\
UnionModel & union\_8 & ((T (T (E) 0 0 (E)) 1 0 (E)),(T (E) 0 1 (E))) \\
UnionPost & union\_8 & ((T (T (E) 0 0 (E)) 1 0 (E)),(T (E) 0 1 (E)),0) \\
UnionUnionAssoc & union\_8 & ((T (E) 0 0 (E)),(T (E) 0 1 (E)),(T (E) -1 0 (E))) \\
\end{longtable}

\begin{longtable}{@{}l l >{\ttfamily\footnotesize}p{0.46\linewidth}@{}}
\caption{Ground-truth minimal counterexamples for RBT (34 of 58 tasks solved by the deterministic LeanCheck search).}\label{tab:gt-rbt}\\
\toprule
\normalfont Property & \normalfont Mutation & \normalfont Minimal counterexample \\
\midrule
\endfirsthead
\multicolumn{3}{@{}l}{\footnotesize\itshape Table~\ref{tab:gt-rbt}, RBT, continued}\\
\toprule
\normalfont Property & \normalfont Mutation & \normalfont Minimal counterexample \\
\midrule
\endhead
\bottomrule
\endlastfoot
DeleteInsert & miscolor\_insert & ((T (B) (E) 1 0 (E)),0,0,0) \\
InsertValid & miscolor\_insert & ((T (B) (E) 0 0 (E)),1,0) \\
DeleteInsert & insert\_1 & ((T (B) (E) 1 0 (E)),0,0,0) \\
InsertInsert & insert\_1 & ((E),0,1,0,0) \\
InsertModel & insert\_1 & ((T (B) (E) 0 0 (E)),1,0) \\
InsertPost & insert\_1 & ((T (B) (E) 0 0 (E)),1,0,0) \\
DeleteInsert & insert\_2 & ((T (B) (E) 0 0 (E)),0,1,0) \\
InsertDelete & insert\_2 & ((T (B) (E) 0 0 (E)),1,0,0) \\
InsertInsert & insert\_2 & ((E),0,1,0,0) \\
InsertModel & insert\_2 & ((T (B) (E) 0 0 (E)),1,0) \\
InsertPost & insert\_2 & ((T (B) (E) 0 0 (E)),1,0,1) \\
InsertDelete & insert\_3 & ((T (B) (E) 0 0 (E)),0,0,1) \\
InsertInsert & insert\_3 & ((E),0,0,0,1) \\
InsertModel & insert\_3 & ((T (B) (E) 0 0 (E)),0,1) \\
InsertPost & insert\_3 & ((T (B) (E) 0 0 (E)),0,0,1) \\
DeleteInsert & no\_balance\_insert\_1 & \normalfont --- \\
InsertDelete & no\_balance\_insert\_1 & \normalfont --- \\
InsertValid & no\_balance\_insert\_1 & ((T (B) (T (R) (E) 0 0 (E)) 1 0 (E)),-1,0) \\
DeleteInsert & no\_balance\_insert\_2 & ((T (B) (E) 0 0 (E)),1,1,0) \\
InsertDelete & no\_balance\_insert\_2 & ((T (B) (E) 0 0 (E)),1,1,0) \\
InsertValid & no\_balance\_insert\_2 & ((T (B) (E) 0 0 (E)),1,0) \\
DeleteValid & miscolor\_delete & ((T (B) (E) 0 0 (E)),1) \\
DeleteDelete & delete\_4 & ((T (B) (T (R) (E) -1 0 (E)) 0 0 (T (R) (E) 1 0 (E))),0,-1) \\
DeleteInsert & delete\_4 & ((E),0,1,0) \\
DeleteModel & delete\_4 & ((T (B) (E) 0 0 (E)),1) \\
DeletePost & delete\_4 & ((T (B) (E) 0 0 (E)),1,0) \\
InsertDelete & delete\_4 & ((E),0,1,0) \\
DeleteDelete & delete\_5 & ((T (B) (E) 0 0 (T (R) (E) 1 0 (E))),0,1) \\
DeleteInsert & delete\_5 & ((T (B) (E) 1 0 (E)),0,0,0) \\
DeleteModel & delete\_5 & ((T (B) (T (R) (E) 0 0 (E)) 1 0 (E)),0) \\
DeletePost & delete\_5 & ((T (B) (T (R) (E) 0 0 (E)) 1 0 (E)),0,0) \\
DeleteDelete & miscolor\_balLeft & \normalfont --- \\
DeleteValid & miscolor\_balLeft & \normalfont --- \\
DeleteDelete & miscolor\_balRight & \normalfont --- \\
DeleteValid & miscolor\_balRight & \normalfont --- \\
DeleteValid & miscolor\_join\_1 & \normalfont --- \\
DeleteDelete & miscolor\_join\_2 & \normalfont --- \\
DeleteValid & miscolor\_join\_2 & \normalfont --- \\
DeleteDelete & swap\_cd & \normalfont --- \\
DeleteInsert & swap\_cd & ((T (B) (T (R) (E) 0 0 (E)) 1 0 (T (R) (E) 2 0 (E))),0,-1,0) \\
DeleteModel & swap\_cd & \normalfont --- \\
DeletePost & swap\_cd & \normalfont --- \\
DeleteValid & swap\_cd & \normalfont --- \\
InsertDelete & swap\_cd & ((T (B) (T (R) (E) 0 0 (E)) 1 0 (T (R) (E) 2 0 (E))),-1,0,0) \\
InsertInsert & swap\_cd & ((T (B) (T (R) (E) 0 0 (E)) 1 0 (E)),-1,2,0,0) \\
InsertModel & swap\_cd & ((T (B) (T (R) (E) 0 0 (E)) 1 0 (T (R) (E) 2 0 (E))),-1,0) \\
InsertPost & swap\_cd & \normalfont --- \\
InsertValid & swap\_cd & ((T (B) (T (R) (E) 0 0 (E)) 1 0 (T (R) (E) 2 0 (E))),-1,0) \\
DeleteDelete & swap\_bc & \normalfont --- \\
DeleteInsert & swap\_bc & \normalfont --- \\
DeleteModel & swap\_bc & \normalfont --- \\
DeletePost & swap\_bc & \normalfont --- \\
DeleteValid & swap\_bc & \normalfont --- \\
InsertDelete & swap\_bc & \normalfont --- \\
InsertInsert & swap\_bc & \normalfont --- \\
InsertModel & swap\_bc & \normalfont --- \\
InsertPost & swap\_bc & \normalfont --- \\
InsertValid & swap\_bc & \normalfont --- \\
\end{longtable}

\begin{longtable}{@{}l l >{\ttfamily\footnotesize}p{0.46\linewidth}@{}}
\caption{Ground-truth minimal counterexamples for STLC (20 of 20 tasks solved by the deterministic LeanCheck search).}\label{tab:gt-stlc}\\
\toprule
\normalfont Property & \normalfont Mutation & \normalfont Minimal counterexample \\
\midrule
\endfirsthead
\multicolumn{3}{@{}l}{\footnotesize\itshape Table~\ref{tab:gt-stlc}, STLC, continued}\\
\toprule
\normalfont Property & \normalfont Mutation & \normalfont Minimal counterexample \\
\midrule
\endhead
\bottomrule
\endlastfoot
MultiPreserve & shift\_var\_none & (Abs (TBool) (App (Abs (TBool) (Var 1)) (Bool \#f))) \\
SinglePreserve & shift\_var\_none & (Abs (TBool) (App (Abs (TBool) (Var 1)) (Bool \#f))) \\
MultiPreserve & shift\_var\_all & (App (Abs (TBool) (Abs (TBool) (Var 0))) (Bool \#f)) \\
SinglePreserve & shift\_var\_all & (App (Abs (TBool) (Abs (TBool) (Var 0))) (Bool \#f)) \\
MultiPreserve & shift\_var\_leq & (Abs (TBool) (App (Abs (TBool) (Abs (TFun(TBool) (TBool)) (Var 1))) (Var 0))) \\
SinglePreserve & shift\_var\_leq & (Abs (TBool) (App (Abs (TBool) (Abs (TFun(TBool) (TBool)) (Var 1))) (Var 0))) \\
MultiPreserve & shift\_abs\_no\_incr & (App (Abs (TBool) (Abs (TBool) (Var 0))) (Bool \#f)) \\
SinglePreserve & shift\_abs\_no\_incr & (App (Abs (TBool) (Abs (TBool) (Var 0))) (Bool \#f)) \\
MultiPreserve & subst\_var\_all & (App (Abs (TBool) (Abs (TFun(TBool) (TBool)) (Var 0))) (Bool \#f)) \\
SinglePreserve & subst\_var\_all & (App (Abs (TBool) (Abs (TFun(TBool) (TBool)) (Var 0))) (Bool \#f)) \\
MultiPreserve & subst\_var\_none & (App (Abs (TBool) (Var 0)) (Bool \#f)) \\
SinglePreserve & subst\_var\_none & (App (Abs (TBool) (Var 0)) (Bool \#f)) \\
MultiPreserve & subst\_abs\_no\_shift & (Abs (TBool) (App (Abs (TBool) (Abs (TFun(TBool) (TBool)) (Var 1))) (Var 0))) \\
SinglePreserve & subst\_abs\_no\_shift & (Abs (TBool) (App (Abs (TBool) (Abs (TFun(TBool) (TBool)) (Var 1))) (Var 0))) \\
MultiPreserve & subst\_abs\_no\_incr & (App (Abs (TBool) (Abs (TFun(TBool) (TBool)) (Var 0))) (Bool \#f)) \\
SinglePreserve & subst\_abs\_no\_incr & (App (Abs (TBool) (Abs (TFun(TBool) (TBool)) (Var 0))) (Bool \#f)) \\
MultiPreserve & substTop\_no\_shift & (Abs (TBool) (App (Abs (TBool) (Var 1)) (Bool \#f))) \\
SinglePreserve & substTop\_no\_shift & (Abs (TBool) (App (Abs (TBool) (Var 1)) (Bool \#f))) \\
MultiPreserve & substTop\_no\_shift\_back & (Abs (TBool) (App (Abs (TBool) (Var 0)) (Var 0))) \\
SinglePreserve & substTop\_no\_shift\_back & (Abs (TBool) (App (Abs (TBool) (Var 0)) (Var 0))) \\
\end{longtable}

\begin{longtable}{@{}l l >{\ttfamily\footnotesize}p{0.46\linewidth}@{}}
\caption{Ground-truth minimal counterexamples for $F_{<:}$ (36 of 36 tasks solved by the deterministic LeanCheck search).}\label{tab:gt-fsub}\\
\toprule
\normalfont Property & \normalfont Mutation & \normalfont Minimal counterexample \\
\midrule
\endfirsthead
\multicolumn{3}{@{}l}{\footnotesize\itshape Table~\ref{tab:gt-fsub}, $F_{<:}$, continued}\\
\toprule
\normalfont Property & \normalfont Mutation & \normalfont Minimal counterexample \\
\midrule
\endhead
\bottomrule
\endlastfoot
MultiPreserve & tshift\_tvar\_all & Abs (All Top (TVar 0)) (TApp (TAbs Top (Var 0)) (All Top Top)) \\
SinglePreserve & tshift\_tvar\_all & Abs (All Top (TVar 0)) (TApp (TAbs Top (Var 0)) (All Top Top)) \\
MultiPreserve & tshift\_tvar\_no\_incr & TApp (TAbs Top (Abs (TVar 0) (TAbs Top (Var 0)))) Top \\
SinglePreserve & tshift\_tvar\_no\_incr & TApp (TAbs Top (Abs (TVar 0) (TAbs Top (Var 0)))) Top \\
MultiPreserve & tshift\_all\_no\_incr & Abs (All Top (TVar 0)) (TApp (TAbs Top (Var 0)) (All Top Top)) \\
SinglePreserve & tshift\_all\_no\_incr & Abs (All Top (TVar 0)) (TApp (TAbs Top (Var 0)) (All Top Top)) \\
MultiPreserve & shift\_var\_all & App (Abs Top (Abs Top (Var 1))) (Abs (All Top Top) (TApp (Var 0) Top)) \\
SinglePreserve & shift\_var\_all & App (Abs Top (Abs Top (Var 1))) (Abs (All Top Top) (TApp (Var 0) Top)) \\
MultiPreserve & shift\_var\_no\_incr & Abs (All Top Top) (App (Abs Top (Abs Top (Var 1))) (TApp (Var 0) Top)) \\
SinglePreserve & shift\_var\_no\_incr & Abs (All Top Top) (App (Abs Top (Abs Top (Var 1))) (TApp (Var 0) Top)) \\
MultiPreserve & shift\_abs\_no\_incr & App (Abs Top (Abs Top (Var 1))) (Abs (All Top Top) (TApp (Var 0) Top)) \\
SinglePreserve & shift\_abs\_no\_incr & App (Abs Top (Abs Top (Var 1))) (Abs (All Top Top) (TApp (Var 0) Top)) \\
MultiPreserve & shift\_typ\_tabs\_no\_incr & App (Abs Top (TAbs Top (Var 0))) (TAbs (All Top Top) (Abs (TVar 0) (TApp (Var 0) Top))) \\
SinglePreserve & shift\_typ\_tabs\_no\_incr & App (Abs Top (TAbs Top (Var 0))) (TAbs (All Top Top) (Abs (TVar 0) (TApp (Var 0) Top))) \\
MultiPreserve & tsubst\_tvar\_flip & Abs (All Top (TVar 0)) (TApp (TAbs Top (Var 0)) Top) \\
SinglePreserve & tsubst\_tvar\_flip & Abs (All Top (TVar 0)) (TApp (TAbs Top (Var 0)) Top) \\
MultiPreserve & tsubst\_tvar\_no\_shift & TAbs Top (Abs (TVar 0) (TApp (TAbs Top (Var 0)) Top)) \\
SinglePreserve & tsubst\_tvar\_no\_shift & TAbs Top (Abs (TVar 0) (TApp (TAbs Top (Var 0)) Top)) \\
MultiPreserve & tsubst\_tvar\_over\_shift & Abs (All Top (TVar 0)) (TApp (TAbs Top (Var 0)) Top) \\
SinglePreserve & tsubst\_tvar\_over\_shift & Abs (All Top (TVar 0)) (TApp (TAbs Top (Var 0)) Top) \\
MultiPreserve & tsubst\_all\_no\_tshift & TAbs Top (TApp (TAbs Top (Abs (TVar 0) (TAbs Top (Var 0)))) (TVar 0)) \\
SinglePreserve & tsubst\_all\_no\_tshift & TAbs Top (TApp (TAbs Top (Abs (TVar 0) (TAbs Top (Var 0)))) (TVar 0)) \\
MultiPreserve & subst\_var\_flip & Abs Top (App (Abs Top (Var 1)) (Var 0)) \\
SinglePreserve & subst\_var\_flip & Abs Top (App (Abs Top (Var 1)) (Var 0)) \\
MultiPreserve & subst\_var\_no\_decr & Abs Top (App (Abs Top (Var 1)) (Var 0)) \\
SinglePreserve & subst\_var\_no\_decr & Abs Top (App (Abs Top (Var 1)) (Var 0)) \\
MultiPreserve & subst\_abs\_no\_shift & Abs (All Top Top) (App (Abs Top (Abs Top (Var 1))) (TApp (Var 0) Top)) \\
SinglePreserve & subst\_abs\_no\_shift & Abs (All Top Top) (App (Abs Top (Abs Top (Var 1))) (TApp (Var 0) Top)) \\
MultiPreserve & subst\_abs\_no\_incr & Abs Top (App (Abs Top (Abs (Arr Top Top) (Var 0))) (Var 0)) \\
SinglePreserve & subst\_abs\_no\_incr & Abs Top (App (Abs Top (Abs (Arr Top Top) (Var 0))) (Var 0)) \\
MultiPreserve & subst\_tabs\_no\_shift & TAbs Top (App (Abs (Arr (TVar 0) Top) (TAbs Top (Var 0))) (Abs (TVar 0) (Var 0))) \\
SinglePreserve & subst\_tabs\_no\_shift & TAbs Top (App (Abs (Arr (TVar 0) Top) (TAbs Top (Var 0))) (Abs (TVar 0) (Var 0))) \\
MultiPreserve & subst\_typ\_tabs\_no\_incr & TApp (TAbs Top (TAbs Top (Abs (TVar 0) (Var 0)))) Top \\
SinglePreserve & subst\_typ\_tabs\_no\_incr & TApp (TAbs Top (TAbs Top (Abs (TVar 0) (Var 0)))) Top \\
MultiPreserve & subst\_typ\_tabs\_no\_shift & TAbs Top (TApp (TAbs Top (TAbs Top (Abs (TVar 1) (Var 0)))) (TVar 0)) \\
SinglePreserve & subst\_typ\_tabs\_no\_shift & TAbs Top (TApp (TAbs Top (TAbs Top (Abs (TVar 1) (Var 0)))) (TVar 0)) \\
\end{longtable}

\section{Full Statistical Comparison}\label{app:stats}

The tables below give, for every (workload, generator family), the Friedman omnibus test across tasks and the post-hoc Holm-corrected pairwise Wilcoxon signed-rank tests, for five metrics: bug-finding time,\footnote{Bug-finding time is the wall-clock time spent before the first failing execution. Tasks for which any of the compared libraries never produces a failing trial are excluded from this metric, so its $N$ can differ from the shrinking metrics.} tree edit distance to the ground-truth minimum,\footnote{Tree edit distance to ground truth, and time per edit, both require the minimal counterexample established by the exhaustive LeanCheck search. This exists for every task of BST, STLC, and $F_{<:}$, but for only 34 of RBT's 58 tasks---the remaining 24 are too deep for exhaustive search. Tasks without a ground truth are excluded from these two metrics, lowering their $N$ relative to shrink time.} counterexample size, shrink time, and time per edit.\footnote{Time per edit ($\mathrm{ms}/\Delta\mathrm{TED}$) is undefined when shrinking produces no reduction in edit distance to the ground truth ($d \le 0$); such tasks are excluded from this metric only, which can lower its $N$ slightly even where ground truth is complete.} Per-task values are trial medians; all metrics are lower-is-better. $\Delta$ is the median per-task difference (first minus second library) and $r$ the matched-pairs rank-biserial effect size (negative $\Rightarrow$ first library better). The reported $N$ is the number of tasks on which all three libraries have a value for that metric.

\begin{longtable}{@{}l l r r r@{}}
\caption{Statistical comparison of bug-finding and shrinking metrics for BST.}\label{tab:stats-bst}\\
\toprule
Comparison & median $\Delta$ & $r$ & $p$ & $p_{\text{Holm}}$ \\
\midrule
\endfirsthead
\multicolumn{5}{@{}l}{\footnotesize\itshape Table~\ref{tab:stats-bst}, BST, continued}\\
\toprule
Comparison & median $\Delta$ & $r$ & $p$ & $p_{\text{Holm}}$ \\
\midrule
\endhead
\bottomrule
\endlastfoot
\multicolumn{5}{@{}l}{\textit{Type-based generators}}\\
\midrule
\multicolumn{5}{@{}l}{\quad \footnotesize Bug-finding time (ms)}\\
\quad Friedman ($N\!=\!48$) & \multicolumn{2}{c}{$\chi^2\!=\!51.5$} & $<\!0.001$ & --- \\
\quad Quick vs.\ Hedgehog & $-56.3$ & $-0.69$ & $<\!0.001$ & $<\!0.001$ \\
\quad Quick vs.\ Falsify & $-13.6$ & $-0.80$ & $<\!0.001$ & $<\!0.001$ \\
\quad Hedgehog vs.\ Falsify & $+16.0$ & $+0.58$ & $<\!0.001$ & $<\!0.001$ \\
\addlinespace
\multicolumn{5}{@{}l}{\quad \footnotesize TED to ground truth}\\
\quad Friedman ($N\!=\!48$) & \multicolumn{2}{c}{$\chi^2\!=\!10.0$} & $0.007$ & --- \\
\quad Quick vs.\ Hedgehog & $+0.0$ & $-0.37$ & $0.105$ & $0.315$ \\
\quad Quick vs.\ Falsify & $+0.0$ & $-0.37$ & $0.253$ & $0.505$ \\
\quad Hedgehog vs.\ Falsify & $+0.0$ & $+0.20$ & $0.372$ & $0.505$ \\
\addlinespace
\multicolumn{5}{@{}l}{\quad \footnotesize Counterexample size}\\
\quad Friedman ($N\!=\!48$) & \multicolumn{2}{c}{$\chi^2\!=\!1.0$} & $0.607$ & --- \\
\quad Quick vs.\ Hedgehog & $+0.0$ & $-1.00$ & $0.317$ & $0.952$ \\
\quad Quick vs.\ Falsify & $+0.0$ & $-1.00$ & $0.317$ & $0.952$ \\
\quad Hedgehog vs.\ Falsify & $+0.0$ & $-0.33$ & $0.655$ & $0.952$ \\
\addlinespace
\multicolumn{5}{@{}l}{\quad \footnotesize Shrink time (ms)}\\
\quad Friedman ($N\!=\!48$) & \multicolumn{2}{c}{$\chi^2\!=\!80.8$} & $<\!0.001$ & --- \\
\quad Quick vs.\ Hedgehog & $-0.2$ & $-0.92$ & $<\!0.001$ & $<\!0.001$ \\
\quad Quick vs.\ Falsify & $-4.0$ & $-0.99$ & $<\!0.001$ & $<\!0.001$ \\
\quad Hedgehog vs.\ Falsify & $-3.9$ & $-0.96$ & $<\!0.001$ & $<\!0.001$ \\
\addlinespace
\multicolumn{5}{@{}l}{\quad \footnotesize Time per edit (ms)}\\
\quad Friedman ($N\!=\!48$) & \multicolumn{2}{c}{$\chi^2\!=\!76.0$} & $<\!0.001$ & --- \\
\quad Quick vs.\ Hedgehog & $-0.1$ & $-1.00$ & $<\!0.001$ & $<\!0.001$ \\
\quad Quick vs.\ Falsify & $-1.3$ & $-0.99$ & $<\!0.001$ & $<\!0.001$ \\
\quad Hedgehog vs.\ Falsify & $-1.2$ & $-0.93$ & $<\!0.001$ & $<\!0.001$ \\
\addlinespace
\midrule
\multicolumn{5}{@{}l}{\textit{API-based generators}}\\
\midrule
\multicolumn{5}{@{}l}{\quad \footnotesize Bug-finding time (ms)}\\
\quad Friedman ($N\!=\!53$) & \multicolumn{2}{c}{$\chi^2\!=\!24.0$} & $<\!0.001$ & --- \\
\quad QuickAPI vs.\ HedgehogAPI & $-2.4$ & $-0.87$ & $<\!0.001$ & $<\!0.001$ \\
\quad QuickAPI vs.\ FalsifyAPI & $-2.3$ & $-0.62$ & $<\!0.001$ & $<\!0.001$ \\
\quad HedgehogAPI vs.\ FalsifyAPI & $+0.1$ & $+0.05$ & $0.740$ & $0.740$ \\
\addlinespace
\multicolumn{5}{@{}l}{\quad \footnotesize TED to ground truth}\\
\quad Friedman ($N\!=\!53$) & \multicolumn{2}{c}{$\chi^2\!=\!12.9$} & $0.002$ & --- \\
\quad QuickAPI vs.\ HedgehogAPI & $+0.0$ & $-0.63$ & $0.006$ & $0.017$ \\
\quad QuickAPI vs.\ FalsifyAPI & $+0.0$ & $+0.01$ & $0.976$ & $0.976$ \\
\quad HedgehogAPI vs.\ FalsifyAPI & $+0.0$ & $+0.43$ & $0.038$ & $0.076$ \\
\addlinespace
\multicolumn{5}{@{}l}{\quad \footnotesize Counterexample size}\\
\quad Friedman ($N\!=\!53$) & \multicolumn{2}{c}{$\chi^2\!=\!0.1$} & $0.949$ & --- \\
\quad QuickAPI vs.\ HedgehogAPI & $+0.0$ & $+0.17$ & $0.785$ & $1.000$ \\
\quad QuickAPI vs.\ FalsifyAPI & $+0.0$ & $+0.29$ & $0.516$ & $1.000$ \\
\quad HedgehogAPI vs.\ FalsifyAPI & $+0.0$ & $+0.20$ & $0.705$ & $1.000$ \\
\addlinespace
\multicolumn{5}{@{}l}{\quad \footnotesize Shrink time (ms)}\\
\quad Friedman ($N\!=\!53$) & \multicolumn{2}{c}{$\chi^2\!=\!100.1$} & $<\!0.001$ & --- \\
\quad QuickAPI vs.\ HedgehogAPI & $-0.5$ & $-1.00$ & $<\!0.001$ & $<\!0.001$ \\
\quad QuickAPI vs.\ FalsifyAPI & $-9.2$ & $-1.00$ & $<\!0.001$ & $<\!0.001$ \\
\quad HedgehogAPI vs.\ FalsifyAPI & $-8.7$ & $-0.99$ & $<\!0.001$ & $<\!0.001$ \\
\addlinespace
\multicolumn{5}{@{}l}{\quad \footnotesize Time per edit (ms)}\\
\quad Friedman ($N\!=\!53$) & \multicolumn{2}{c}{$\chi^2\!=\!34.7$} & $<\!0.001$ & --- \\
\quad QuickAPI vs.\ HedgehogAPI & $-0.0$ & $-0.93$ & $<\!0.001$ & $<\!0.001$ \\
\quad QuickAPI vs.\ FalsifyAPI & $-0.1$ & $-0.77$ & $<\!0.001$ & $<\!0.001$ \\
\quad HedgehogAPI vs.\ FalsifyAPI & $-0.1$ & $-0.38$ & $0.017$ & $0.017$ \\
\addlinespace
\midrule
\multicolumn{5}{@{}l}{\textit{Correct-by-construction generators}}\\
\midrule
\multicolumn{5}{@{}l}{\quad \footnotesize Bug-finding time (ms)}\\
\quad Friedman ($N\!=\!53$) & \multicolumn{2}{c}{$\chi^2\!=\!1.8$} & $0.397$ & --- \\
\quad QuickCBC vs.\ HedgehogCBC & $-0.0$ & $+0.04$ & $0.794$ & $0.794$ \\
\quad QuickCBC vs.\ FalsifyCBC & $-4.2$ & $-0.61$ & $<\!0.001$ & $<\!0.001$ \\
\quad HedgehogCBC vs.\ FalsifyCBC & $-4.2$ & $-0.61$ & $<\!0.001$ & $<\!0.001$ \\
\addlinespace
\multicolumn{5}{@{}l}{\quad \footnotesize TED to ground truth}\\
\quad Friedman ($N\!=\!53$) & \multicolumn{2}{c}{$\chi^2\!=\!54.2$} & $<\!0.001$ & --- \\
\quad QuickCBC vs.\ HedgehogCBC & $-5.0$ & $-0.96$ & $<\!0.001$ & $<\!0.001$ \\
\quad QuickCBC vs.\ FalsifyCBC & $-3.0$ & $-0.98$ & $<\!0.001$ & $<\!0.001$ \\
\quad HedgehogCBC vs.\ FalsifyCBC & $+0.0$ & $-0.33$ & $0.066$ & $0.066$ \\
\addlinespace
\multicolumn{5}{@{}l}{\quad \footnotesize Counterexample size}\\
\quad Friedman ($N\!=\!53$) & \multicolumn{2}{c}{$\chi^2\!=\!42.1$} & $<\!0.001$ & --- \\
\quad QuickCBC vs.\ HedgehogCBC & $-6.0$ & $-1.00$ & $<\!0.001$ & $<\!0.001$ \\
\quad QuickCBC vs.\ FalsifyCBC & $+0.0$ & $-1.00$ & $<\!0.001$ & $<\!0.001$ \\
\quad HedgehogCBC vs.\ FalsifyCBC & $+0.0$ & $-0.12$ & $0.539$ & $0.539$ \\
\addlinespace
\multicolumn{5}{@{}l}{\quad \footnotesize Shrink time (ms)}\\
\quad Friedman ($N\!=\!53$) & \multicolumn{2}{c}{$\chi^2\!=\!81.6$} & $<\!0.001$ & --- \\
\quad QuickCBC vs.\ HedgehogCBC & $-0.5$ & $-0.87$ & $<\!0.001$ & $<\!0.001$ \\
\quad QuickCBC vs.\ FalsifyCBC & $-15.6$ & $-1.00$ & $<\!0.001$ & $<\!0.001$ \\
\quad HedgehogCBC vs.\ FalsifyCBC & $-14.6$ & $-0.89$ & $<\!0.001$ & $<\!0.001$ \\
\addlinespace
\multicolumn{5}{@{}l}{\quad \footnotesize Time per edit (ms)}\\
\quad Friedman ($N\!=\!53$) & \multicolumn{2}{c}{$\chi^2\!=\!72.1$} & $<\!0.001$ & --- \\
\quad QuickCBC vs.\ HedgehogCBC & $-0.0$ & $-0.97$ & $<\!0.001$ & $<\!0.001$ \\
\quad QuickCBC vs.\ FalsifyCBC & $-0.2$ & $-0.97$ & $<\!0.001$ & $<\!0.001$ \\
\quad HedgehogCBC vs.\ FalsifyCBC & $-0.1$ & $-0.76$ & $<\!0.001$ & $<\!0.001$ \\
\addlinespace
\end{longtable}

\begin{longtable}{@{}l l r r r@{}}
\caption{Statistical comparison of bug-finding and shrinking metrics for RBT.}\label{tab:stats-rbt}\\
\toprule
Comparison & median $\Delta$ & $r$ & $p$ & $p_{\text{Holm}}$ \\
\midrule
\endfirsthead
\multicolumn{5}{@{}l}{\footnotesize\itshape Table~\ref{tab:stats-rbt}, RBT, continued}\\
\toprule
Comparison & median $\Delta$ & $r$ & $p$ & $p_{\text{Holm}}$ \\
\midrule
\endhead
\bottomrule
\endlastfoot
\multicolumn{5}{@{}l}{\textit{Type-based generators}}\\
\midrule
\multicolumn{5}{@{}l}{\quad \footnotesize Bug-finding time (ms)}\\
\quad Friedman ($N\!=\!28$) & \multicolumn{2}{c}{$\chi^2\!=\!38.8$} & $<\!0.001$ & --- \\
\quad Quick vs.\ Hedgehog & $-173.1$ & $-0.86$ & $<\!0.001$ & $<\!0.001$ \\
\quad Quick vs.\ Falsify & $-16.5$ & $-0.74$ & $<\!0.001$ & $<\!0.001$ \\
\quad Hedgehog vs.\ Falsify & $+76.7$ & $+0.86$ & $<\!0.001$ & $<\!0.001$ \\
\addlinespace
\multicolumn{5}{@{}l}{\quad \footnotesize TED to ground truth}\\
\quad Friedman ($N\!=\!28$) & \multicolumn{2}{c}{$\chi^2\!=\!6.6$} & $0.038$ & --- \\
\quad Quick vs.\ Hedgehog & $+0.0$ & $-0.41$ & $0.205$ & $0.615$ \\
\quad Quick vs.\ Falsify & $+0.0$ & $+0.04$ & $0.932$ & $1.000$ \\
\quad Hedgehog vs.\ Falsify & $+0.0$ & $+0.17$ & $0.607$ & $1.000$ \\
\addlinespace
\multicolumn{5}{@{}l}{\quad \footnotesize Counterexample size}\\
\quad Friedman ($N\!=\!28$) & \multicolumn{2}{c}{$\chi^2\!=\!0.0$} & $1.000$ & --- \\
\quad Quick vs.\ Hedgehog & $+0.0$ & $+0.00$ & $1.000$ & $1.000$ \\
\quad Quick vs.\ Falsify & $+0.0$ & $+0.00$ & $1.000$ & $1.000$ \\
\quad Hedgehog vs.\ Falsify & $+0.0$ & $+0.00$ & $1.000$ & $1.000$ \\
\addlinespace
\multicolumn{5}{@{}l}{\quad \footnotesize Shrink time (ms)}\\
\quad Friedman ($N\!=\!28$) & \multicolumn{2}{c}{$\chi^2\!=\!52.3$} & $<\!0.001$ & --- \\
\quad Quick vs.\ Hedgehog & $-0.3$ & $-1.00$ & $<\!0.001$ & $<\!0.001$ \\
\quad Quick vs.\ Falsify & $-2.8$ & $-1.00$ & $<\!0.001$ & $<\!0.001$ \\
\quad Hedgehog vs.\ Falsify & $-2.5$ & $-0.98$ & $<\!0.001$ & $<\!0.001$ \\
\addlinespace
\multicolumn{5}{@{}l}{\quad \footnotesize Time per edit (ms)}\\
\quad Friedman ($N\!=\!28$) & \multicolumn{2}{c}{$\chi^2\!=\!46.5$} & $<\!0.001$ & --- \\
\quad Quick vs.\ Hedgehog & $-0.2$ & $-1.00$ & $<\!0.001$ & $<\!0.001$ \\
\quad Quick vs.\ Falsify & $-1.5$ & $-1.00$ & $<\!0.001$ & $<\!0.001$ \\
\quad Hedgehog vs.\ Falsify & $-1.3$ & $-0.95$ & $<\!0.001$ & $<\!0.001$ \\
\addlinespace
\midrule
\multicolumn{5}{@{}l}{\textit{API-based generators}}\\
\midrule
\multicolumn{5}{@{}l}{\quad \footnotesize Bug-finding time (ms)}\\
\quad Friedman ($N\!=\!56$) & \multicolumn{2}{c}{$\chi^2\!=\!54.3$} & $<\!0.001$ & --- \\
\quad QuickAPI vs.\ HedgehogAPI & $-32.2$ & $-0.97$ & $<\!0.001$ & $<\!0.001$ \\
\quad QuickAPI vs.\ FalsifyAPI & $-42.6$ & $-0.84$ & $<\!0.001$ & $<\!0.001$ \\
\quad HedgehogAPI vs.\ FalsifyAPI & $+0.3$ & $+0.29$ & $0.060$ & $0.060$ \\
\addlinespace
\multicolumn{5}{@{}l}{\quad \footnotesize TED to ground truth}\\
\quad Friedman ($N\!=\!34$) & \multicolumn{2}{c}{$\chi^2\!=\!1.8$} & $0.415$ & --- \\
\quad QuickAPI vs.\ HedgehogAPI & $+0.0$ & $+0.44$ & $0.070$ & $0.139$ \\
\quad QuickAPI vs.\ FalsifyAPI & $+0.0$ & $+0.61$ & $0.026$ & $0.077$ \\
\quad HedgehogAPI vs.\ FalsifyAPI & $+0.0$ & $-0.28$ & $0.284$ & $0.284$ \\
\addlinespace
\multicolumn{5}{@{}l}{\quad \footnotesize Counterexample size}\\
\quad Friedman ($N\!=\!56$) & \multicolumn{2}{c}{$\chi^2\!=\!13.5$} & $0.001$ & --- \\
\quad QuickAPI vs.\ HedgehogAPI & $+0.0$ & $+0.41$ & $0.146$ & $0.146$ \\
\quad QuickAPI vs.\ FalsifyAPI & $+0.0$ & $-0.43$ & $0.066$ & $0.133$ \\
\quad HedgehogAPI vs.\ FalsifyAPI & $+0.0$ & $-0.89$ & $<\!0.001$ & $0.002$ \\
\addlinespace
\multicolumn{5}{@{}l}{\quad \footnotesize Shrink time (ms)}\\
\quad Friedman ($N\!=\!56$) & \multicolumn{2}{c}{$\chi^2\!=\!110.0$} & $<\!0.001$ & --- \\
\quad QuickAPI vs.\ HedgehogAPI & $-1.4$ & $-1.00$ & $<\!0.001$ & $<\!0.001$ \\
\quad QuickAPI vs.\ FalsifyAPI & $-48.9$ & $-1.00$ & $<\!0.001$ & $<\!0.001$ \\
\quad HedgehogAPI vs.\ FalsifyAPI & $-47.9$ & $-0.99$ & $<\!0.001$ & $<\!0.001$ \\
\addlinespace
\multicolumn{5}{@{}l}{\quad \footnotesize Time per edit (ms)}\\
\quad Friedman ($N\!=\!34$) & \multicolumn{2}{c}{$\chi^2\!=\!25.9$} & $<\!0.001$ & --- \\
\quad QuickAPI vs.\ HedgehogAPI & $-0.1$ & $-0.99$ & $<\!0.001$ & $<\!0.001$ \\
\quad QuickAPI vs.\ FalsifyAPI & $-0.0$ & $-0.71$ & $<\!0.001$ & $<\!0.001$ \\
\quad HedgehogAPI vs.\ FalsifyAPI & $-0.0$ & $-0.18$ & $0.379$ & $0.379$ \\
\addlinespace
\midrule
\multicolumn{5}{@{}l}{\textit{Correct-by-construction generators}}\\
\midrule
\multicolumn{5}{@{}l}{\quad \footnotesize Bug-finding time (ms)}\\
\quad Friedman ($N\!=\!42$) & \multicolumn{2}{c}{$\chi^2\!=\!54.1$} & $<\!0.001$ & --- \\
\quad QuickCBC vs.\ HedgehogCBC & $-60.1$ & $-0.98$ & $<\!0.001$ & $<\!0.001$ \\
\quad QuickCBC vs.\ FalsifyCBC & $-26.0$ & $-0.88$ & $<\!0.001$ & $<\!0.001$ \\
\quad HedgehogCBC vs.\ FalsifyCBC & $+42.6$ & $+0.94$ & $<\!0.001$ & $<\!0.001$ \\
\addlinespace
\multicolumn{5}{@{}l}{\quad \footnotesize TED to ground truth}\\
\quad Friedman ($N\!=\!30$) & \multicolumn{2}{c}{$\chi^2\!=\!21.4$} & $<\!0.001$ & --- \\
\quad QuickCBC vs.\ HedgehogCBC & $-0.5$ & $-0.57$ & $0.025$ & $0.038$ \\
\quad QuickCBC vs.\ FalsifyCBC & $-1.5$ & $-0.80$ & $<\!0.001$ & $0.002$ \\
\quad HedgehogCBC vs.\ FalsifyCBC & $-1.0$ & $-0.57$ & $0.019$ & $0.038$ \\
\addlinespace
\multicolumn{5}{@{}l}{\quad \footnotesize Counterexample size}\\
\quad Friedman ($N\!=\!42$) & \multicolumn{2}{c}{$\chi^2\!=\!9.7$} & $0.008$ & --- \\
\quad QuickCBC vs.\ HedgehogCBC & $+0.0$ & $-0.55$ & $0.078$ & $0.156$ \\
\quad QuickCBC vs.\ FalsifyCBC & $+0.0$ & $-0.82$ & $0.015$ & $0.045$ \\
\quad HedgehogCBC vs.\ FalsifyCBC & $+0.0$ & $-0.15$ & $0.633$ & $0.633$ \\
\addlinespace
\multicolumn{5}{@{}l}{\quad \footnotesize Shrink time (ms)}\\
\quad Friedman ($N\!=\!42$) & \multicolumn{2}{c}{$\chi^2\!=\!34.5$} & $<\!0.001$ & --- \\
\quad QuickCBC vs.\ HedgehogCBC & $-0.4$ & $+0.12$ & $0.516$ & $0.516$ \\
\quad QuickCBC vs.\ FalsifyCBC & $-5.0$ & $-0.94$ & $<\!0.001$ & $<\!0.001$ \\
\quad HedgehogCBC vs.\ FalsifyCBC & $-5.0$ & $-0.88$ & $<\!0.001$ & $<\!0.001$ \\
\addlinespace
\multicolumn{5}{@{}l}{\quad \footnotesize Time per edit (ms)}\\
\quad Friedman ($N\!=\!30$) & \multicolumn{2}{c}{$\chi^2\!=\!39.5$} & $<\!0.001$ & --- \\
\quad QuickCBC vs.\ HedgehogCBC & $-0.2$ & $-0.88$ & $<\!0.001$ & $<\!0.001$ \\
\quad QuickCBC vs.\ FalsifyCBC & $-0.1$ & $-0.99$ & $<\!0.001$ & $<\!0.001$ \\
\quad HedgehogCBC vs.\ FalsifyCBC & $+0.0$ & $-0.26$ & $0.213$ & $0.213$ \\
\addlinespace
\end{longtable}

\begin{longtable}{@{}l l r r r@{}}
\caption{Statistical comparison of bug-finding and shrinking metrics for STLC.}\label{tab:stats-stlc}\\
\toprule
Comparison & median $\Delta$ & $r$ & $p$ & $p_{\text{Holm}}$ \\
\midrule
\endfirsthead
\multicolumn{5}{@{}l}{\footnotesize\itshape Table~\ref{tab:stats-stlc}, STLC, continued}\\
\toprule
Comparison & median $\Delta$ & $r$ & $p$ & $p_{\text{Holm}}$ \\
\midrule
\endhead
\bottomrule
\endlastfoot
\multicolumn{5}{@{}l}{\textit{Type-based generators}}\\
\midrule
\multicolumn{5}{@{}l}{\quad \footnotesize Bug-finding time (ms)}\\
\quad Friedman ($N\!=\!16$) & \multicolumn{2}{c}{$\chi^2\!=\!1.6$} & $0.444$ & --- \\
\quad Quick vs.\ Hedgehog & $-7.8$ & $-0.16$ & $0.597$ & $1.000$ \\
\quad Quick vs.\ Falsify & $+47.6$ & $+0.13$ & $0.669$ & $1.000$ \\
\quad Hedgehog vs.\ Falsify & $+143.6$ & $+0.19$ & $0.528$ & $1.000$ \\
\addlinespace
\multicolumn{5}{@{}l}{\quad \footnotesize TED to ground truth}\\
\quad Friedman ($N\!=\!16$) & \multicolumn{2}{c}{$\chi^2\!=\!14.5$} & $<\!0.001$ & --- \\
\quad Quick vs.\ Hedgehog & $-2.5$ & $-0.78$ & $0.006$ & $0.017$ \\
\quad Quick vs.\ Falsify & $-1.0$ & $-0.56$ & $0.058$ & $0.117$ \\
\quad Hedgehog vs.\ Falsify & $+1.0$ & $+0.54$ & $0.086$ & $0.117$ \\
\addlinespace
\multicolumn{5}{@{}l}{\quad \footnotesize Counterexample size}\\
\quad Friedman ($N\!=\!16$) & \multicolumn{2}{c}{$\chi^2\!=\!8.2$} & $0.017$ & --- \\
\quad Quick vs.\ Hedgehog & $-2.0$ & $-0.88$ & $0.008$ & $0.024$ \\
\quad Quick vs.\ Falsify & $+0.0$ & $-0.79$ & $0.054$ & $0.107$ \\
\quad Hedgehog vs.\ Falsify & $+0.0$ & $+0.53$ & $0.133$ & $0.133$ \\
\addlinespace
\multicolumn{5}{@{}l}{\quad \footnotesize Shrink time (ms)}\\
\quad Friedman ($N\!=\!16$) & \multicolumn{2}{c}{$\chi^2\!=\!32.0$} & $<\!0.001$ & --- \\
\quad Quick vs.\ Hedgehog & $-0.3$ & $-1.00$ & $<\!0.001$ & $<\!0.001$ \\
\quad Quick vs.\ Falsify & $-23.5$ & $-1.00$ & $<\!0.001$ & $<\!0.001$ \\
\quad Hedgehog vs.\ Falsify & $-23.3$ & $-1.00$ & $<\!0.001$ & $<\!0.001$ \\
\addlinespace
\multicolumn{5}{@{}l}{\quad \footnotesize Time per edit (ms)}\\
\quad Friedman ($N\!=\!16$) & \multicolumn{2}{c}{$\chi^2\!=\!25.1$} & $<\!0.001$ & --- \\
\quad Quick vs.\ Hedgehog & $-0.0$ & $-0.35$ & $0.231$ & $0.231$ \\
\quad Quick vs.\ Falsify & $-2.1$ & $-1.00$ & $<\!0.001$ & $<\!0.001$ \\
\quad Hedgehog vs.\ Falsify & $-2.1$ & $-1.00$ & $<\!0.001$ & $<\!0.001$ \\
\addlinespace
\midrule
\multicolumn{5}{@{}l}{\textit{Correct-by-construction generators}}\\
\midrule
\multicolumn{5}{@{}l}{\quad \footnotesize Bug-finding time (ms)}\\
\quad Friedman ($N\!=\!20$) & \multicolumn{2}{c}{$\chi^2\!=\!24.1$} & $<\!0.001$ & --- \\
\quad QuickCBC vs.\ HedgehogCBC & $-4.1$ & $-0.99$ & $<\!0.001$ & $<\!0.001$ \\
\quad QuickCBC vs.\ FalsifyCBC & $-1.5$ & $-0.94$ & $<\!0.001$ & $<\!0.001$ \\
\quad HedgehogCBC vs.\ FalsifyCBC & $+0.9$ & $+0.42$ & $0.105$ & $0.105$ \\
\addlinespace
\multicolumn{5}{@{}l}{\quad \footnotesize TED to ground truth}\\
\quad Friedman ($N\!=\!20$) & \multicolumn{2}{c}{$\chi^2\!=\!14.7$} & $<\!0.001$ & --- \\
\quad QuickCBC vs.\ HedgehogCBC & $-4.5$ & $+0.00$ & $1.000$ & $1.000$ \\
\quad QuickCBC vs.\ FalsifyCBC & $+5.0$ & $+0.54$ & $0.033$ & $0.066$ \\
\quad HedgehogCBC vs.\ FalsifyCBC & $+7.8$ & $+1.00$ & $<\!0.001$ & $<\!0.001$ \\
\addlinespace
\multicolumn{5}{@{}l}{\quad \footnotesize Counterexample size}\\
\quad Friedman ($N\!=\!20$) & \multicolumn{2}{c}{$\chi^2\!=\!14.7$} & $<\!0.001$ & --- \\
\quad QuickCBC vs.\ HedgehogCBC & $-4.2$ & $+0.01$ & $0.968$ & $0.968$ \\
\quad QuickCBC vs.\ FalsifyCBC & $+5.8$ & $+0.61$ & $0.022$ & $0.044$ \\
\quad HedgehogCBC vs.\ FalsifyCBC & $+7.8$ & $+1.00$ & $<\!0.001$ & $<\!0.001$ \\
\addlinespace
\multicolumn{5}{@{}l}{\quad \footnotesize Shrink time (ms)}\\
\quad Friedman ($N\!=\!20$) & \multicolumn{2}{c}{$\chi^2\!=\!40.0$} & $<\!0.001$ & --- \\
\quad QuickCBC vs.\ HedgehogCBC & $-0.6$ & $-1.00$ & $<\!0.001$ & $<\!0.001$ \\
\quad QuickCBC vs.\ FalsifyCBC & $-6.5$ & $-1.00$ & $<\!0.001$ & $<\!0.001$ \\
\quad HedgehogCBC vs.\ FalsifyCBC & $-5.9$ & $-1.00$ & $<\!0.001$ & $<\!0.001$ \\
\addlinespace
\multicolumn{5}{@{}l}{\quad \footnotesize Time per edit (ms)}\\
\quad Friedman ($N\!=\!20$) & \multicolumn{2}{c}{$\chi^2\!=\!36.4$} & $<\!0.001$ & --- \\
\quad QuickCBC vs.\ HedgehogCBC & $-0.0$ & $-0.95$ & $<\!0.001$ & $<\!0.001$ \\
\quad QuickCBC vs.\ FalsifyCBC & $-0.3$ & $-1.00$ & $<\!0.001$ & $<\!0.001$ \\
\quad HedgehogCBC vs.\ FalsifyCBC & $-0.3$ & $-1.00$ & $<\!0.001$ & $<\!0.001$ \\
\addlinespace
\end{longtable}

\begin{longtable}{@{}l l r r r@{}}
\caption{Statistical comparison of bug-finding and shrinking metrics for $F_{<:}$.}\label{tab:stats-fsub}\\
\toprule
Comparison & median $\Delta$ & $r$ & $p$ & $p_{\text{Holm}}$ \\
\midrule
\endfirsthead
\multicolumn{5}{@{}l}{\footnotesize\itshape Table~\ref{tab:stats-fsub}, $F_{<:}$, continued}\\
\toprule
Comparison & median $\Delta$ & $r$ & $p$ & $p_{\text{Holm}}$ \\
\midrule
\endhead
\bottomrule
\endlastfoot
\multicolumn{5}{@{}l}{\textit{Type-based generators}}\\
\midrule
\multicolumn{5}{@{}l}{\quad \footnotesize Bug-finding time (ms)}\\
\quad Friedman ($N\!=\!26$) & \multicolumn{2}{c}{$\chi^2\!=\!46.7$} & $<\!0.001$ & --- \\
\quad Quick vs.\ Hedgehog & $-1055.2$ & $-1.00$ & $<\!0.001$ & $<\!0.001$ \\
\quad Quick vs.\ Falsify & $-115.4$ & $-0.89$ & $<\!0.001$ & $<\!0.001$ \\
\quad Hedgehog vs.\ Falsify & $+776.8$ & $+1.00$ & $<\!0.001$ & $<\!0.001$ \\
\addlinespace
\multicolumn{5}{@{}l}{\quad \footnotesize TED to ground truth}\\
\quad Friedman ($N\!=\!26$) & \multicolumn{2}{c}{$\chi^2\!=\!12.5$} & $0.002$ & --- \\
\quad Quick vs.\ Hedgehog & $-2.5$ & $-0.88$ & $<\!0.001$ & $0.001$ \\
\quad Quick vs.\ Falsify & $-2.2$ & $-0.81$ & $0.001$ & $0.002$ \\
\quad Hedgehog vs.\ Falsify & $+0.0$ & $+0.45$ & $0.079$ & $0.079$ \\
\addlinespace
\multicolumn{5}{@{}l}{\quad \footnotesize Counterexample size}\\
\quad Friedman ($N\!=\!26$) & \multicolumn{2}{c}{$\chi^2\!=\!29.5$} & $<\!0.001$ & --- \\
\quad Quick vs.\ Hedgehog & $-3.0$ & $-0.96$ & $<\!0.001$ & $<\!0.001$ \\
\quad Quick vs.\ Falsify & $-2.2$ & $-0.94$ & $<\!0.001$ & $0.001$ \\
\quad Hedgehog vs.\ Falsify & $+0.5$ & $+0.88$ & $0.002$ & $0.002$ \\
\addlinespace
\multicolumn{5}{@{}l}{\quad \footnotesize Shrink time (ms)}\\
\quad Friedman ($N\!=\!26$) & \multicolumn{2}{c}{$\chi^2\!=\!52.0$} & $<\!0.001$ & --- \\
\quad Quick vs.\ Hedgehog & $-0.2$ & $-1.00$ & $<\!0.001$ & $<\!0.001$ \\
\quad Quick vs.\ Falsify & $-2.7$ & $-1.00$ & $<\!0.001$ & $<\!0.001$ \\
\quad Hedgehog vs.\ Falsify & $-2.5$ & $-1.00$ & $<\!0.001$ & $<\!0.001$ \\
\addlinespace
\multicolumn{5}{@{}l}{\quad \footnotesize Time per edit (ms)}\\
\quad Friedman ($N\!=\!26$) & \multicolumn{2}{c}{$\chi^2\!=\!48.3$} & $<\!0.001$ & --- \\
\quad Quick vs.\ Hedgehog & $-0.0$ & $-0.91$ & $<\!0.001$ & $<\!0.001$ \\
\quad Quick vs.\ Falsify & $-0.3$ & $-1.00$ & $<\!0.001$ & $<\!0.001$ \\
\quad Hedgehog vs.\ Falsify & $-0.2$ & $-1.00$ & $<\!0.001$ & $<\!0.001$ \\
\addlinespace
\midrule
\multicolumn{5}{@{}l}{\textit{Correct-by-construction generators}}\\
\midrule
\multicolumn{5}{@{}l}{\quad \footnotesize Bug-finding time (ms)}\\
\quad Friedman ($N\!=\!36$) & \multicolumn{2}{c}{$\chi^2\!=\!14.9$} & $<\!0.001$ & --- \\
\quad QuickCBC vs.\ HedgehogCBC & $-2.1$ & $-0.75$ & $<\!0.001$ & $<\!0.001$ \\
\quad QuickCBC vs.\ FalsifyCBC & $-0.7$ & $-0.59$ & $0.001$ & $0.003$ \\
\quad HedgehogCBC vs.\ FalsifyCBC & $+0.4$ & $+0.49$ & $0.010$ & $0.010$ \\
\addlinespace
\multicolumn{5}{@{}l}{\quad \footnotesize TED to ground truth}\\
\quad Friedman ($N\!=\!36$) & \multicolumn{2}{c}{$\chi^2\!=\!56.0$} & $<\!0.001$ & --- \\
\quad QuickCBC vs.\ HedgehogCBC & $-60.2$ & $-1.00$ & $<\!0.001$ & $<\!0.001$ \\
\quad QuickCBC vs.\ FalsifyCBC & $-52.0$ & $-1.00$ & $<\!0.001$ & $<\!0.001$ \\
\quad HedgehogCBC vs.\ FalsifyCBC & $+7.2$ & $+0.40$ & $0.035$ & $0.035$ \\
\addlinespace
\multicolumn{5}{@{}l}{\quad \footnotesize Counterexample size}\\
\quad Friedman ($N\!=\!36$) & \multicolumn{2}{c}{$\chi^2\!=\!56.0$} & $<\!0.001$ & --- \\
\quad QuickCBC vs.\ HedgehogCBC & $-46.0$ & $-1.00$ & $<\!0.001$ & $<\!0.001$ \\
\quad QuickCBC vs.\ FalsifyCBC & $-39.5$ & $-1.00$ & $<\!0.001$ & $<\!0.001$ \\
\quad HedgehogCBC vs.\ FalsifyCBC & $+5.2$ & $+0.39$ & $0.039$ & $0.039$ \\
\addlinespace
\multicolumn{5}{@{}l}{\quad \footnotesize Shrink time (ms)}\\
\quad Friedman ($N\!=\!36$) & \multicolumn{2}{c}{$\chi^2\!=\!62.0$} & $<\!0.001$ & --- \\
\quad QuickCBC vs.\ HedgehogCBC & $-1.6$ & $-1.00$ & $<\!0.001$ & $<\!0.001$ \\
\quad QuickCBC vs.\ FalsifyCBC & $-21.8$ & $-1.00$ & $<\!0.001$ & $<\!0.001$ \\
\quad HedgehogCBC vs.\ FalsifyCBC & $-19.7$ & $-0.92$ & $<\!0.001$ & $<\!0.001$ \\
\addlinespace
\multicolumn{5}{@{}l}{\quad \footnotesize Time per edit (ms)}\\
\quad Friedman ($N\!=\!36$) & \multicolumn{2}{c}{$\chi^2\!=\!68.2$} & $<\!0.001$ & --- \\
\quad QuickCBC vs.\ HedgehogCBC & $-0.0$ & $-1.00$ & $<\!0.001$ & $<\!0.001$ \\
\quad QuickCBC vs.\ FalsifyCBC & $-0.2$ & $-1.00$ & $<\!0.001$ & $<\!0.001$ \\
\quad HedgehogCBC vs.\ FalsifyCBC & $-0.2$ & $-0.95$ & $<\!0.001$ & $<\!0.001$ \\
\addlinespace
\end{longtable}

\bibliographystyle{ACM-Reference-Format}
\bibliography{references}


\end{document}